\documentclass[preprint, 12pt]{elsarticle}
\usepackage{amsmath}
\usepackage{amsfonts}   
\usepackage{amssymb} 
\usepackage{enumerate}
\usepackage{wrapfig}
\usepackage{fancyhdr}
\usepackage{acronym}
\usepackage{setspace}
\usepackage{fancyvrb}
\def\bA{\mbox{\bf A}}

\def\bH{\mbox{\bf H}}
\def\bI{\mbox{\bf I}}
\def\bK{\mbox{\bf K}}

\def\bQ{\mbox{\bf Q}}
\def\bR{\mbox{\bf R}}

\def\bY{\mbox{\bf Y}}

\def\bu{\mbox{\bf u}}
\def\bx{\mbox{\bf x}}
\def\by{\mbox{\bf y}}

\def\br{\mbox{\bf r}}

\acrodef{EnGSF}{ensemble Gaussian sum filter}
\acrodef{pdf}{probability distribution function}
\acrodef{KF}{Kalman filter}
\acrodef{EnKF}{ensemble Kalman filter}
\acrodef{EnSRF}{ensemble square root filter}
\acrodef{EKF}{extended Kalman filter}
\acrodef{GPR}{Ground-Penetrating Radar}
\acrodef{ERT}{Electrical Resistivity Tomography}
\acrodef{EM}{Electromagnetic}
\acrodef{MC}{Monte Carlo}
\acrodef{GN}{Gauss-Newton}
\acrodef{LM}{Levenberg-Marquardt}
\acrodef{PDE}{Partial Differential Equation}
\acrodef{SVD}{Singular Value Decomposition}
\acrodef{MCMC}{Markov chain Monte Carlo}
\acrodef{GA}{Genetic Algorithm}
\acrodef{SNR}{signal-to-noise ratio}
\acrodef{FMM}{Fast Multipole Method}
\acrodef{FFT}{Fast Fourier Transform}
\acrodef{FGT}{Fast Gauss Transform}
\acrodef{IFGT}{Improved Fast Gauss Transform}
\acrodef{SIS}{Sequential Importance Sampling}
\acrodef{RMSE}{Root Mean Square Error}
\acrodef{SDE}{stochastic differential equation}
\acrodef{SIR}{Sequential Importance Resampling}

\begin{document}

\begin{frontmatter}

\title{An Improved Data Assimilation Scheme for High Dimensional Nonlinear Systems}




\author{Hatef Monajemi \corref{cor1}}
\author{Peter Kitanidis \fnref{label2}}

\fntext[label2]{Professor, Dept. of Civil \& Environmental Eng., Stanford University, USA, Email: peterk@stanford.edu}

\cortext[cor1]{Corresponding Author. Stanford University, Dept. of Civil \& Environmental Eng., Yang \& Yamazaki Environment \& Energy Building, 473 Via Ortega, Stanford, CA 94305, USA, Tel: +1 (650) 391-6374, Fax: + 1 (650) 725-9720,  Email: monajemi@stanford.edu }

\begin{abstract}
Nonlinear/non-Gaussian filtering has broad applications in many areas of life sciences where either the dynamic is nonlinear and/or the probability density function of uncertain state is non-Gaussian. In such problems, the accuracy of the estimated quantities depends highly upon how accurately their posterior pdf can be approximated. In low dimensional state spaces, methods based on Sequential Importance Sampling (SIS) can suitably approximate the posterior pdf. For higher dimensional problems, however, these techniques are usually inappropriate since the required number of particles to achieve satisfactory estimates grows exponentially with the dimension of state space. On the other hand, ensemble Kalman filter (EnKF) and its variants are more suitable for large-scale problems due to transformation of particles in the Bayesian update step. It has been shown that the latter class of methods may lead to suboptimal solutions for strongly nonlinear problems due to the Gaussian assumption in the update step. In this paper, we introduce a new technique based on the Gaussian sum expansion which captures the non-Gaussian features more accurately while the required computational effort remains within reason for high dimensional problems. We demonstrate the performance of the method for non-Gaussian processes through several examples including the strongly nonlinear Lorenz models. Results show a remarkable improvement in the mean square error compared to EnKF, and a desirable convergence behavior as the number of particles increases.
\end{abstract}


\end{frontmatter}

{D}{ata} assimilation is an essential tool for reliable prediction of natural and physical phenomena. It has broad application in many fields of science and engineering such as flood forecasting, weather prediction, contaminant tracking, wild fire tracking, oil exploration, etc. \cite{kitanidis80, mandel09, evensen09, subber08, bishop01}. It involves incorporating sparse observational data (usually corrupted by noise) into computer models to obtain reliable estimates of unknown state and parameters of a dynamical system.

Many data assimilation techniques have been suggested to date that usually fall into one of the two main categories: {\itshape variational} and {\itshape sequential}. Variational methods deal with batch of data at specific time interval and aim to find the best possible estimate by minimizing a penalty function which usually includes the model, initial and boundary condition, and observational errors defined under certain statistical hypothesis \cite{belanger05}. Examples of this kind include 4D-var \cite{talagrand87} and representer method \cite{bennett92}. On the other hand, in sequential data assimilation, which is also the approach adopted in this paper, the model variables are updated every time observations become available and may thus be computationally less expensive than the former \cite{evensen06}. Examples of such techniques include Kalman filter and its variants \cite{kalman60, evensen03, bishop01}.   

In recent years, there has been a significant interest in developing sequential filtering strategies that could more accurately predict the state of a nonlinear process when the number of unknown variables is considerable. van Leeuwen \cite{vanleeuwen09} gives an excellent review of recently developed nonlinear filtering schemes. These methods are mostly based on a Monte Carlo approximation of \ac{pdf}'s of concern. 

There exists \ac{SIS} based filters that represent a \ac{pdf} as the weighted sum of delta functions centered at finite number of particles. In the simplest form, the weights are updated upon receiving observations while the particles remain intact \cite{doucet00}. Such approach usually requires the user of \ac{SIS} to carefully choose a proposal density function from which the particles are drawn since it is desirable to have particles in areas with high likelihood. 

There are also \ac{EnKF} based algorithms that approximate a \ac{pdf} using i.i.d. samples (i.e., equal weights). Upon receiving observations, the weights remain equal while the samples (particles) are moved. \ac{EnKF} was proposed by Evensen \cite{evensen94, burgers98} and has been successfully applied to many large-scale nonlinear estimation problems so far \cite{evensen03, houtekamer98, evensen09, monajemi09}. Much of the success of \ac{EnKF} over SIS filter for large-scale problems is attributed to the fact that it nudges the particles to areas of higher likelihood \cite{vanleeuwen09}. Moreover, it is attractive in that it is conceptually simple, easy to implement and does not require computing the adjoint operator as needed in variational setting. Despite all its success in the nonlinear estimation arena, its tendency to generate unimodal pdf's may result in suboptimal results when the posterior \ac{pdf} is far from Gaussian distribution \cite{mandel09,vanleeuwen09}. This is mainly attributed to the fact that in the Bayesian update step, \ac{EnKF} approximates the prior density function with a Gaussian \ac{pdf} \cite{evensen03}. 

As computers become more powerful, resolution and complexity of numerical models tend to increase resulting in strongly nonlinear systems and hence there is a need to develop new strategies that could more accurately capture the non-Gaussian statistics \cite{vanleeuwen09, vanleeuwen10}. It is this fact that led to the development of compound filters that try to combine the \ac{EnKF} with \ac{SIS} in order to obtain a more accurate representation of posterior \ac{pdf}. Examples of such attempts can be found in \cite{mandel09, vanleeuwen10}. The idea behind these methods is to use the posterior pdf obtained from the \ac{EnKF} as the proposal density for the \ac{SIS} filter. Though these methods are attractive in that they approximate the posterior pdf more accurately than the \ac{EnKF} does, in the opinion of the author, the two stage filtering introduces new tuning parameters (c.f. \cite{mandel09}) and imposes extra computational cost as a result of explicit density kernel estimation which makes it hard to justify for large dimensions. The above mentioned difficulties are what motivated this study to reexamine the fundamentals of ensemble filtering and to possibly address some of these issues.

In what follows we introduce a new methodology, the ensemble Gaussian sum filter (EnGSF) which tends to approximate the non-Gaussian statistics more accurately. This paper is organized as follows: In section \ref{sec:PS} we introduce the general filtering problem. The theory and details of the \ac{EnGSF} is presented in section \ref{sec:EnGSF}. Section \ref{sec:NI} contains several numerical examples that shows the performance the methodology for nonlinear estimation problems. Finally, conclusion and future work are presented in section \ref{sec:concl}.

\section{Problem Statement} \label{sec:PS}
Filtering in control and estimation theory refers to a process comprised of two stages, one of forecast and the other of data assimilation by which one can anticipate the dynamics of natural and physical systems. To further clarify, let us assume that the following Morkov model represents the dynamics of the system under study 
\begin{equation}
\bx_{k+1} = f(\bx_{k}) + {{\bf \epsilon}_{k}}
\label{eq:model}
\end{equation}
where $f(\cdot): R^m \rightarrow R^m $ is the so-called model operator, $\bx_k \in R^m$ is the state of the system at time $k$ and ${\bf \epsilon} \in R^m$ is the noise component representing the modeling uncertainty. Such equation is usually derived from discretization of a set of partial differential equations.

Moreover, suppose that observational data are available at distinct instances in time and are related to the uncertain state variables through the following equation:

\begin{equation}
\by_{k} = \bH \bx_{k} + {{\bf \br}_{k}}
\label{eq:obs}
\end{equation} 
Here, $\by_k \in R^n$ is the vector of observations at time $k$, $\bH_{n \times m}$ is the measurement operator and $\br \in R^n$ is the measurement noise vector.

The uncertain state variable $\bx_{k}$ can be fully described by a \ac{pdf}. The assimilation step of filtering then involves exploiting the observational data to improve the knowledge on the \ac{pdf} of the process while the forecast step refers to evolving the state variables using model equation \ref{eq:model}. 

In terms of \ac{pdf}'s, the forecast step involves finding \textit{prior} \ac{pdf} at time $k$ using information up to time $k-1$, $p(\bx_{k}|\by_{1:k-1})$, through Chapman-Kolmogorov equation \cite{arulampalam02}
\begin{equation}
p(\bx_{k}|\by_{1:k-1}) = \int p(\bx_{k}|\bx_{k-1}) \ p(\bx_{k-1}|\by_{1:k-1}) \ d\bx_{k-1}
\label{eq:CK}
\end{equation}
The assimilation step involves obtaining the \textit{posterior} pdf which contains information up to time $k$, $ p(\bx_{k}|\by_{1:k}) $, using Baye's theorm
\begin{equation}
p(\bx_{k}|\by_{1:k}) = \frac{p(\by_k | \bx_{k}) \ p(\bx_{k}|\by_{1:k-1})}{\int p(\by_k | \bx_{k}) \ p(\bx_{k}|\by_{1:k-1})  \  d\bx_k}
\label{eq:bayes}  
\end{equation}
 
The resulting \ac{pdf}'s in equations (\ref{eq:CK}) and (\ref{eq:bayes}) can not in general be determined analytically unless for linear dynamics under certain assumptions \cite{kalman60, jazwinski07}. For linear dynamics in which only Gaussian \ac{pdf}'s are involved, an optimal filter exists which is known as \ac{KF} \cite{kalman60} in the relevant literature. For nonlinear/non-Gaussian cases, however, there is practically no optimal filter. This has led to the development of numerous sub-optimal filters that seek an estimate either by linearizing the nonlinear model (e.g. \ac{EKF}) or approximating the \ac{pdf} of the process by a collection of particles (e.g. \ac{EnKF}, \ac{SIS} filter)\cite{vanleeuwen09, evensen03, arulampalam02}.

\section{Ensemble Gaussian Sum Filter} \label{sec:EnGSF}
A non-Gaussian \ac{pdf} can be approximated as the sum of finite number of Gaussian kernels known as Gaussian sum expansion\cite{aoki65, sorenson71}. To the best of author's knowledge, the early attempt to use such expansion for nonlinear filtering problems is the work of Sorenson in early 70's \cite{sorenson71}. It is desirable since the sum of Gaussian kernels is always a valid density function regardless of the number of terms used in the expansion and converges uniformly to any desired density function \cite{sorenson71}. 

In this section, we first present the fundamentals of Gaussian sum expansion and subsequently introduce an improved filtering methodology for large-scale nonlinear/non-Gaussian dynamics. We call the new method, the ensemble Gaussian sum filter (EnGSF) since it is built upon the ideas of \textit{ensemble} data assimilation and principles of \textit{Gaussian sum filter}.

\subsection{Gaussian sum expansion}

Since the Gaussian kernel is considered amongst the \textit{delta family of positive} functions, any arbitrary pdf $g$ can be approximated by $g_{\Sigma}$ as \cite{sorenson71}, 
\begin{equation}
g_{\Sigma}(\bx) = \int_{-\infty}^{\infty} N(\bx - \bu, \Sigma) \ g(\bu) \  d\bu
\label{eq:dfp}
\end{equation}
$g_{\Sigma}$ converges uniformly to $g$ as $\Sigma \rightarrow \bf{0}$. Therefore, a discrete approximation to $g(\bx)$ may be obtained as follows:

\begin{equation}
g(\bx) \approx \sum_{i = 1}^{l} \alpha_i \  N(\bx - \bu_i , \Sigma_i)
\label{eq:gs}
\end{equation}      
where \[\alpha_i = g(\bu_i) \  \Delta\bu_i\]
we refer to $\alpha_i$'s as the weights of Gaussian kernels throughout this paper. It is notable as the variance $\Sigma$ decreases, the Gaussian kernels become closer to \textit{Dirac delta} functions and the approximation in the limit of zero variance become identical to the one used in the \ac{SIS} filter.  
\subsection{Bayesian update}
In this section, the Baye's update equation, eq. (\ref{eq:bayes}), is approximated using the Gaussian sum expansion as the basis of our approximation. In the following derivation, we assume that the measurement noise in eq. (\ref{eq:obs}) is distributed according to $N(\textbf{0},\bR)$. We also drop the time subscript $k$ and denote the prior pdf by $p(\bx)$ to avoid complication in notation. Suppose that prior \ac{pdf} is given by:

\begin{equation}
p(\bx) = \sum_{i = 1}^{l} \alpha_i^f \  N(\bx - \bx_i^f , \Sigma_i^f)
\end{equation}      
One can then show that the posterior pdf is given by \cite{sorenson71}:
\begin{equation}
p(\bx|\by) = \sum_{i = 1}^{l} \alpha_i^a \  N(\bx - \bx_i^a , \Sigma_i^a)
\end{equation}    
where 
\begin{equation}
\alpha_i^a = \frac{\alpha_i^f \ N\left(\by - \bH \bx_i^f , \left[\bH \Sigma_i^f \bH^T + \bR\right] \right)} 
{\sum_{i = 1}^l \alpha_i^f \ N\left(\by - \bH \bx_i^f , \left[\bH \Sigma_i^f \bH^T + \bR\right] \right)}
\label{eq:ana_weight}
\end{equation}    
\\
\begin{equation}
\bx_i^a = \bx_i^f + \Sigma_i^f \bH^T \left[\bH \Sigma_i^f \bH^T + \bR \right]^{-1} (\by - \bH \bx_i^f)
\label{eq:ana_mean}
\end{equation}    
\\
\begin{equation}
\Sigma_i^a = \left(\bI - \Sigma_i^f \bH^T \left[\bH \Sigma_i^f \bH^T + \bR \right]^{-1} \bH\right) \Sigma_i^f
\label{eq:ana_cov}
\end{equation}

Throughout this paper, we use superscripts \textit{``f"} and \textit{``a''} to represent \textit{forecast} (prior to the receipt of data) and \textit{analysis} (after assimilation of data) respectively.

According to the above formulas, a choice of $(\alpha_i^f, \bx_i^f, \Sigma_i^f )$ uniquely determines $(\alpha_i^a, \bx_i^a, \Sigma_i^a)$. Therefore, an appropriate choice of prior parameters is crucial for obtaining an accurate estimate of the posterior pdf. 

The weights and particle supports, $(\alpha_i^f, \bx_i^f)$, can be assigned upon initialization of the filter whereas the covariance matrices $\Sigma_i$'s will be calculated at each assimilation step from the \textit{statistics of the ensemble of particles}. Therefore, it is convenient to start filter with i.i.d samples of the prior. It is noted that sampling from the prior when the number of variables is large is not a trivial task. However, for Gaussian \ac{pdf}'s with special covariance structure, there are methods based on \ac{FFT} \cite{dietrich97, wood94} or polynomial expansion of the square root of covariance matrix \cite{dietrich95} by which one can generate initial realizations in a reasonable period of time. The fundamental assumption here is that the initial setup of the filter will be forgotten after forward integration of a dynamical system in a certain period of time.   

Moreover, it is both practical and convenient to choose the same covariance matrix for all particles. Such assumption significantly reduces the cost of data assimilation. In what follows, we give the theory for choosing an appropriate covariance matrix.

\subsubsection{A good choice of $\Sigma^f$}
Choosing an appropriate covariance matrix for individual Gaussian kernels is a crucial step since it ultimately determines the direction and amount of movement for each individual particle. This is usually discussed under the rubric of \textit{bandwidth} selection in density estimation literature \cite{silverman86}.  

$\Sigma^f$ can be found by minimizing the Mean Integrated Squared Error (MISE) between the true prior pdf, $p(\bx)$, and the one estimated using Gaussian sum, $g_{\Sigma}(\bx)$ \cite{scott05, silverman86}:

\begin{eqnarray}   
\text{MISE}\{g_{\Sigma}(\bx)\} &=& E \left \lbrace {\int \left[p(\bx) - g_{\Sigma}(\bx) \right]^2} \ d\bx \right \rbrace  \nonumber \\
 & = & \int {[E\{g_{\Sigma}(\bx)\} - p(\bx)]^2} dx  + \int \text{var} \ g_{\Sigma}(\bx) dx \nonumber \\
 & = & \int {\text{bias}^2} \ dx  + \int \text{var} \ dx
\end{eqnarray}
where $E\{\cdot\}$ indicates the theoretical expectation. Here we eliminated superscript ``$f$" for the sake of clarity. According to the above equations, the problem of determining the optimal choice of $\Sigma$ is a trade-off between bias in the approximation of a pdf and variance of the probability weights. In addition, the optimization procedure requires the complete knowledge of the prior probability which is usually not available. This makes the optimal choice to some extent subjective though it is possible to find approximate solutions for it. It is shown in \cite{silverman86} that assuming a Gaussian pdf as the underlying density at this stage is an appropriate assumption in many cases. Based on this assumption an approximate covariance matrix is obtained as \cite{silverman86}:
\begin{equation}
\Sigma = c N^{-\frac{2}{m+4}} P
\end{equation} 
where $N$ is the number of particles, $m$ is the dimension of state vector, $P$ is the theoretical prior covariance and $c$ is a constanct defined by:
\begin{equation}
c = \left({\frac{4}{m+2}}\right)^{\frac{2}{m+4}}
\end{equation} 
Note that $c$ varies between 0.85 and 1.12 with an assymptotic value of 1.0 for large $m$. Therefore it can be safely replaced by $1.0$, leading to:
\begin{equation}
\Sigma =  N^{-\frac{2}{m+4}} P
\label{eq:srt}
\end{equation}  
The above equations indicate that $\Sigma$ depends upon two factors: (a) the number of particles and (b) the dimension of state space. Given a fixed number of particles, as the dimension of state space increases, larger covariance matrices needed for individual particles to efficiently approximate prior pdf. Note that equation (\ref{eq:srt}) is sometimes referred to as the \textit{Silverman's rule of thumb} and is only optimal for Gaussian pdfs. It tends to oversmoothen the non-Gaussian densities and hence it is usually beneficial to use a smaller covariance matrix in practice as suggested also in \cite{silverman86}. In this paper, we adopt a slightly modified equation for the covariance of individual samples which is given by:
\begin{equation}
\Sigma^f =  N^{-\frac{2}{m+2}} P
\label{eq:sigma}
\end{equation}  
where we have changed the power of $N$ such that the covariance is always smaller than that obtained by equation (\ref{eq:srt}). It is worth noting that the expression for $\Sigma^f$ given by equation (\ref{eq:sigma}) is consistent with the bandwidth suggested by Scott \cite{scott05} based on the theory of histograms and tend to generate more satisfactory estimates in lower dimensions according to authors' experience. For higher dimensional problems, the effect of this change is minimal as $\Sigma^f \rightarrow P$.

At this point, inspired by the ensemble data assimilation strategies, we replace the true (theoretical) covariance P by the weighted ensemble (empirical) covariance arriving at the following equation for $\Sigma^f$:

\begin{equation}
\Sigma^f =  N^{-\frac{2}{m+2}} P_e
\label{eq:sigma_final}
\end{equation}  
where the weighted ensemble covariance matrix is defined by: 
\begin{equation}
P_e = \sum_i \alpha_i (\bx_i - \bar{\bx})  (\bx_i - \bar{\bx})^T
\end{equation}
where $\bar{\bx} = \sum_i \alpha_i  \bx_i $ is the weighted empirical mean. Eq. (\ref{eq:sigma_final}) is the expression used in all the numerical examples performed in this paper.

\subsection{Forecast Step} 
To predict the state of a system in the absence of data, one needs to calculate the Chapman-Kolmogrov integral equation,
\begin{eqnarray*}
p(\bx_{k}|\by_{1:k-1}) &=& \int p(\bx_{k}|\bx_{k-1}) \ p(\bx_{k-1}|\by_{1:k-1}) \ d\bx_{k-1} \\
&=& \sum_{i=1}^l \alpha_i N\left(\bx^k- f(\bx_i^{k-1}), \bQ_i \right) 
\end{eqnarray*}
where $\bQ_i$ is the covariance of each sample after integration. In \ac{EKF} type algoithms, this covariance is calculated using derivatives of the nonlinear function $f(\bx)$ at the current estimate. This is the approach adopted in \cite{sorenson71} for example. In this paper, however, we replace $\bQ_i$ with $\Sigma^f$ given by equation (\ref{eq:sigma_final}) derived from the statistics of the forecast ensemble.  
Based on the the above formula, calculating the integral only involves forward integration of the model equation (\ref{eq:model}) for individual particles.  

\subsection{Degeneracy and Resampling}

Since the methodology discussed in this paper assumes that the weights associated with particles may be different, it is expected that over time these weights become inconsistently non-uniform and after certain time a few particles carry most of the probability mass. This phenomenon which is named \textit{degeneracy} in filtering literature is a common attribute of almost all particle filtering techniques \cite{vanleeuwen09}. For the SIS particle filter, it is shown that the variance of the weights can only increase over time\cite{doucet00} and hence degeneracy is unavoidable. Degeneracy is undesirable since one makes extra effort updating particles that are of minimal weight in addition to the issues that may arise in approximating the covariance when needed. Moreover, for the times when there is no data (i.e., prediction steps) a large noise in the forward integration of the stochastic dynamical model for particles of high weight may take those to locations far from the true state resulting in an undesirable forecasts whereas this phenomenon is usually not present in case with equally probable particles. 

To reduce the effect of degeneracy, one usually resamples the posterior \ac{pdf} when the number of particles with considerable weight decreases. Resampling generally involves ignoring particles with small weights and focusing on the particles that carry larger weights. The reader if referred to \cite{doucet00} for a detailed discussion on different resampling strategies. An appropriate measure of degeneracy is the \textit{effective} number of particles, $N_{eff}$ introduced in \cite{kong94} and is defined by,

\begin{equation}
N_{eff} = \frac{N}{1 + var(\alpha^*)}
\end{equation}        

where $\alpha^*$ is the so-called true weight. This expression cannot be evaluated exactly, however, it can be estimated using the following formula \cite{doucet00, arulampalam02}:

\begin{equation}
\hat{N}_{eff} = \frac{1}{\sum_{i =1}^{N} \alpha_i^2}
\end{equation}      

In SIS filter, one usually resamples if $N_{eff}$ becomes less than a given treshold, $N_{treshold}$. In our approach, we prefer to resample every time data are assimilated since \ac{EnGSF} relies on the covariance matrix obtained from the particles. A typical resampling algorithm is given below:

\begin{Verbatim}[frame = topline]

Algorithm 1: Resample
\end{Verbatim}
$[\{\bx_{out}^j, \alpha_{out}^j\}_{j=1}^N ] = \verb+Resample+[\{\bx_{in}^i, \alpha_{in}^i\}_{i=1}^N]$ 
\begin{itemize}
\item \verb+Initialize CDF:+ $c_0 = 0$
\item \verb+FOR + $i = 1: N$ \\
- \verb+Construct CDF:+ $c_i = c_{i-1} + \alpha_{in}^i$ 
\item \verb+END FOR+ 
\item \verb+Draw N random number+ $\{u^j\}_{j = 1}^N $\verb+according to+ $U[0,1]$ 
\item \verb+Start at the bottom of CDF: + $i = 0$
\item \verb+FOR + $j = 1: N$ \\
- \verb+WHILE + $c_i < u^j$\\
 * $i = i+1$ \\
- \verb+END WHILE+ \\
- \verb+Assign sample:+ $\bx_{out}^j = \bx_{in}^i $ \\
- \verb+Assign weight:+ $\alpha_{out}^j = N^{-1}$ 
\item \verb+END FOR+
\end{itemize}
\begin{Verbatim}[frame = bottomline]

\end{Verbatim}

Note that the weighting step of EnGSF is quite similar to that of SIS, however, particles are moved to the best possible positions after weighting unlike the SIS filter in which particles are fixed in their positions. Transformation of particles after weighting is what prevents filter divergence even in high-dimensional cases with small number of particles. 

From another perspective, EnGSF may be looked as a collection of EnKFs acting together and are weighted according to their importance. The importance is measured by how close a specific particle (i.e., the mean of a Gaussian kernel) is to the measurement with respect to a suitable distance measure.      

It is a known fact that resampling may reduce the diversity of particles and hence modeling uncertainty is an essential tool in such settings to introduce more diversity. Note that in the \ac{EnGSF}, each analysis particle carries a Gaussian kernel which may be used to introduce more diversity. In particular, when only one particle carry all the probability mass (i.e., $\alpha_s^a = 1.0$ for $\bx_s^a$), we can perform a Gaussian re-sampling step by the following transformation:  

\begin{eqnarray}
\bA^a_{m\times N} =  \left[ \bx_s^a, \bx_s^a,\cdots, \bx_s^a \right]_{m\times N}+ \left\{ \bA'^f + \bK (\bY' - \bH \bA'^f) \right\} _{m \times N}
\label{eq:gr}
\end{eqnarray}
where 
\begin{eqnarray}
\bK =  \Sigma^f \bH^T \left( \bH \Sigma^f \bH^T + \bR \right)^{-1},
\end{eqnarray}
\begin{eqnarray}
\Sigma^f =  \bA'^f  {\bA'^f}^T 
\end{eqnarray}
and columns of the so-called measurement and state perturbation matrices are given by: 
\begin{eqnarray}
{\bA'^f}_{(:,j)} = \left( \alpha_j \ N^{\frac{- 2}{m+2}} \right)^{\frac{1}{2}} \ (\bx^f_{j} - \bar{\bx^f})
\end{eqnarray}
\begin{eqnarray}
\bY'_{(:,j)}  = \by + \br_j,   \ \ \ \br_j \sim N(\textbf{0}, \bR)
\end{eqnarray}

The above equations are the same introduced in \cite{burgers98}. It can be shown that samples obtained using eq. (\ref{eq:gr}) will asymptotically (i.e., in the limit of very large ensemble) have the desired mean and covariance given by eqs. (\ref{eq:ana_mean}) and (\ref{eq:ana_cov}) respectively. Thus Gaussian resampling involves adding adequate perturbations to the analysis particles to achieve the desired spread. Though not explored in this work, it may be beneficial to use the resampling ideas from the recently advised Merging Particle Filter \cite{nakano07}.

\subsection{The complete algorithm for \ac{EnGSF}} \label{subsec:engsf_algo}
The complete algorithm for the \ac{EnGSF} is summarized below:

\begin{Verbatim}[frame = topline]

Algorithm 2: EnGSF 
\end{Verbatim}
$[\{\bx_i^k, \alpha_i^k\}_{i=1}^N ] = \verb+EnGSF+[\{\bx_i^{k-1}, \alpha_i^{k-1}\}_{i=1}^N, \by^k]$ 
\begin{itemize}
\item \verb+FOR + $i = 1: N$ \\
- \verb+Integrate according to forward model+ (\ref{eq:model}) \\
$[\{\bx_i^f, \alpha_i^f\}_{i=1}^N ] = \verb+fwd_model+ [\{\bx_i^{k-1}, \alpha_i^{k-1}\}_{i=1}^N]$
\item \verb+END FOR+ 
\item \verb+FOR + $i = 1: N$ \\
- \verb+Calculate weights + $\alpha_i^a$ \verb+ according to+ eq. (\ref{eq:ana_weight}) \\
- \verb+Calculate analysis particles + $\bx_i^a$ \verb+ according to+ eq. (\ref{eq:ana_mean})
\item \verb+END FOR+
\item \verb+Resample using algorithm 1+ \\
 - $[\{\bx_i^k, \alpha_i^k\}_{i=1}^N ] = \verb+Resample+[\{\bx_i^{a}, \alpha_i^{a}\}_{i=1}^N]$ 
\end{itemize}
\begin{Verbatim}[frame = bottomline]

\end{Verbatim}

\section{Numerical Illustration} \label{sec:NI}
In this section, we provide four numerical examples to test the performance of our method in forecasting state of nonlinear dynamics. The first two examples are one dimensional and have previously appeared in \cite{mandel09}. The third and forth examples are the celebrated three and forty dimensional Lorenz models \cite{lorenz63, lorenz95} which have been studied extensively in the data assimilation literature.    
\subsection{Example 1: One-dimensional Bayesian update}
In this example, we study one step of a non-Gaussian Bayesian update. We consider prior and likelihood pdf's of the following forms
 
\begin{equation}
p(x) = \frac{ 1} {\eta} \left[ e^{- \frac{1}{2} (\frac{x -1.5}{0.1})^2} +   e^{- \frac{1}{2} (\frac{x+1.5}{0.1})^2} \right]
\end{equation} 
\begin{equation}
p(d|x) = \frac{1 } {\sqrt{2 \pi (0.1)} }  e^{- \frac{1}{2} (\frac{d-x}{0.1})^2}
\end{equation} 

\begin{figure}[htbp]
\centering
\includegraphics[width= 200pt,keepaspectratio]{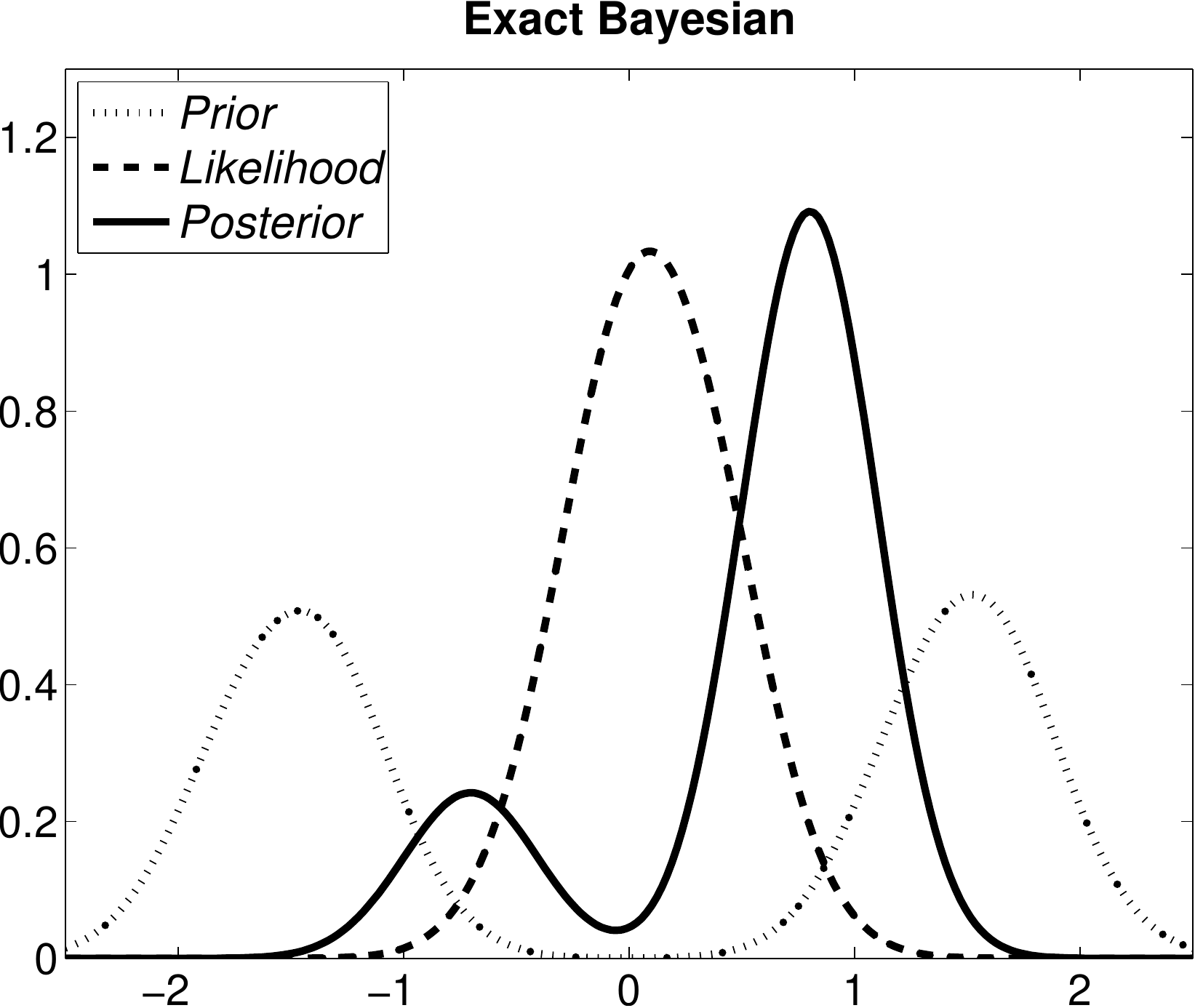}
\caption{Prior, likelihood and the true posterior pdf obtained by direct multiplication of prior and likelihood at $10000$ uniform grid points in $[-4, \ 4]$}
\label{fig:exact_posterior}
\end{figure}

Here $\eta$ is a normalization constant so that $\int_{-\infty}^{\infty} p(x) \ dx = 1$. Such set up may happen in nonlinear dynamics when the system is in transition between stable points as will be demonstrated in the next example. We draw i.i.d samples from the prior and estimate the posterior pdf using different methods for various number of particles. The true pdf's are plotted in Fig.~ (\ref{fig:exact_posterior}).

\begin{figure}[htbp]
\begin{tabular}{p{-0pt}|ccc}

\begin{minipage}{15pt}
\centering

\end{minipage}
&
\begin{minipage}{150pt}
\centering
$N = 200$
\end{minipage}
&
\hspace{-50pt}
\begin{minipage}{150pt}
\centering
$N = 500$
\end{minipage}
&
\hspace{-50pt}
\begin{minipage}{150pt}
\centering
$N = 1000$
\end{minipage}
\vspace{10pt}

\\
\hline

\\
\begin{minipage}{15pt}
\centering
\hspace{-50pt}
EnKF
\end{minipage}
&
\begin{minipage}{125pt}
\centering
\includegraphics[width=125pt]{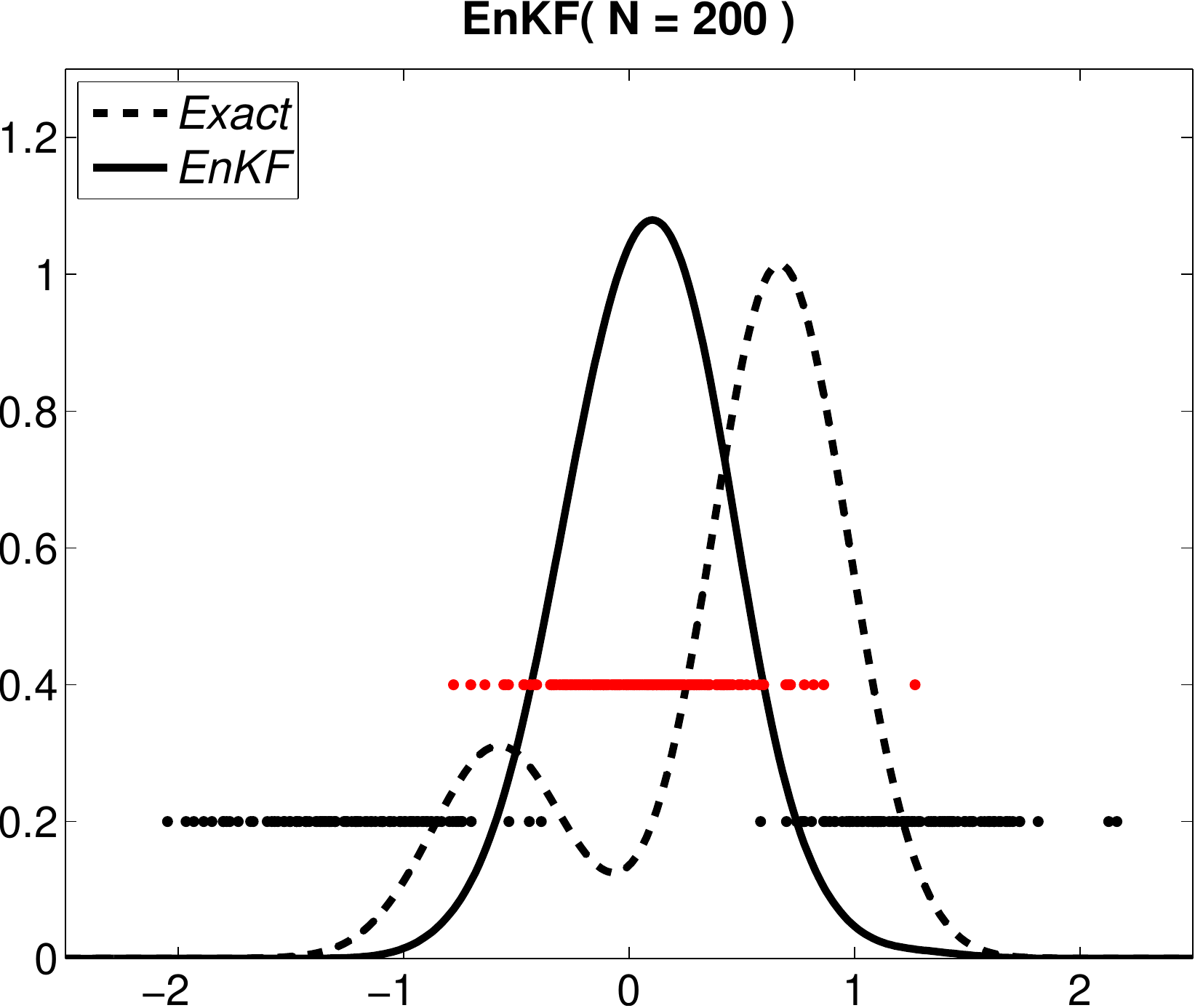}
\end{minipage}
&
\hspace{-25pt}
\begin{minipage}{125pt}
\centering
\includegraphics[width=125pt]{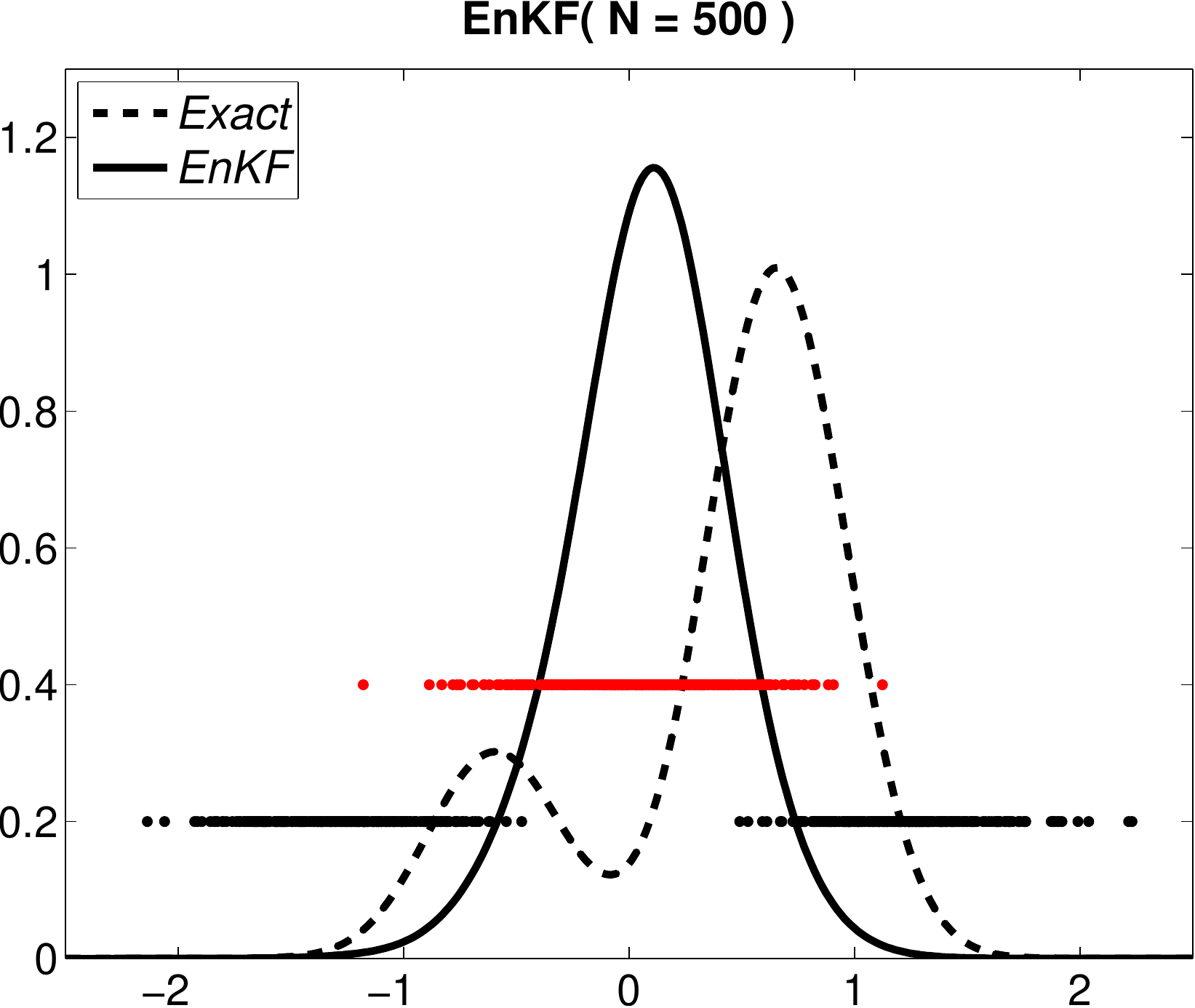}
\end{minipage}
&
\hspace{-25pt}
\begin{minipage}{125pt}
\centering
\includegraphics[width=125pt]{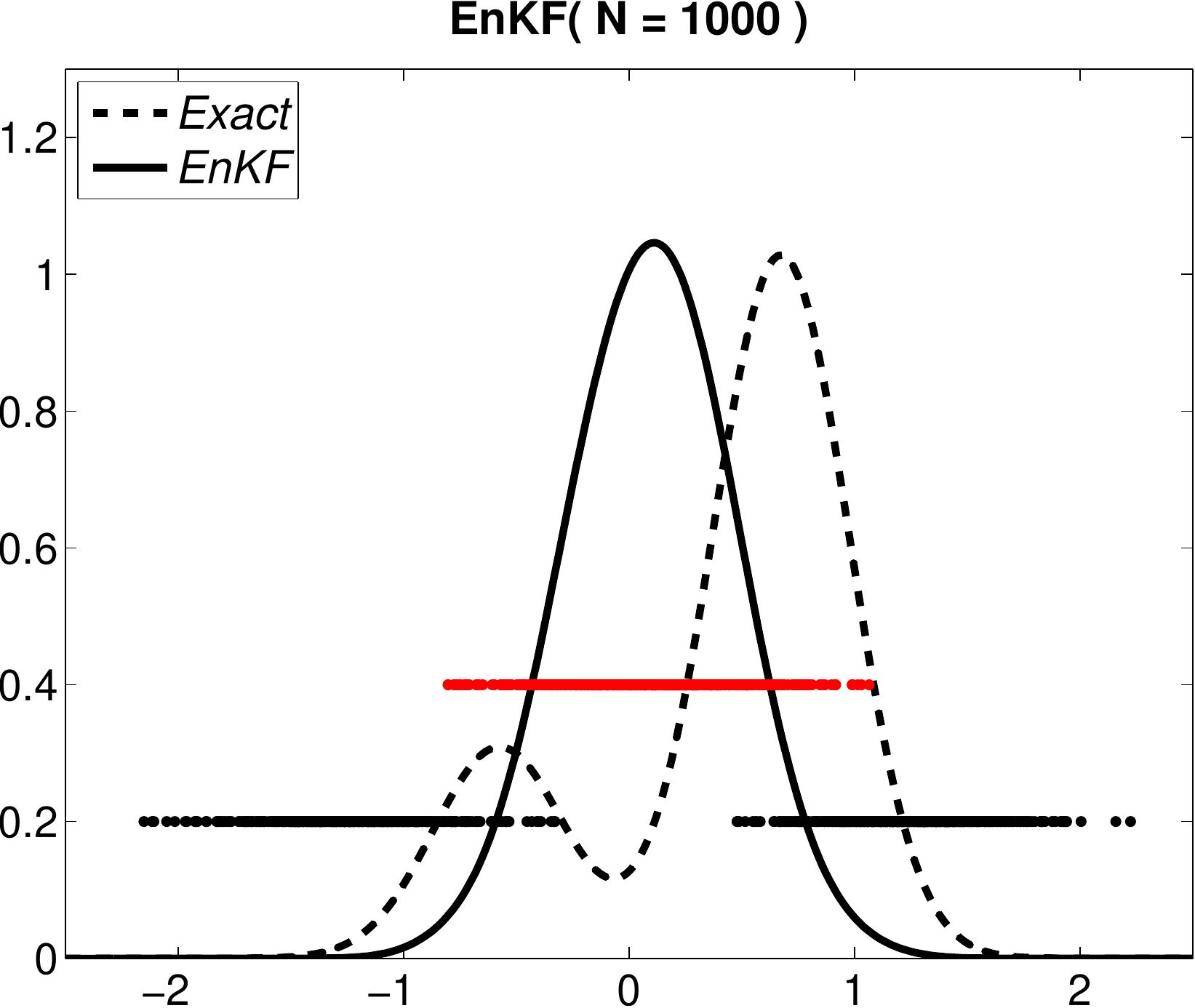}
\end{minipage}

\\
\begin{minipage}{15pt}
\centering
\hspace{-50pt}
EnSRF
\end{minipage}
&
\begin{minipage}{125pt}
\centering
\includegraphics[width=125pt]{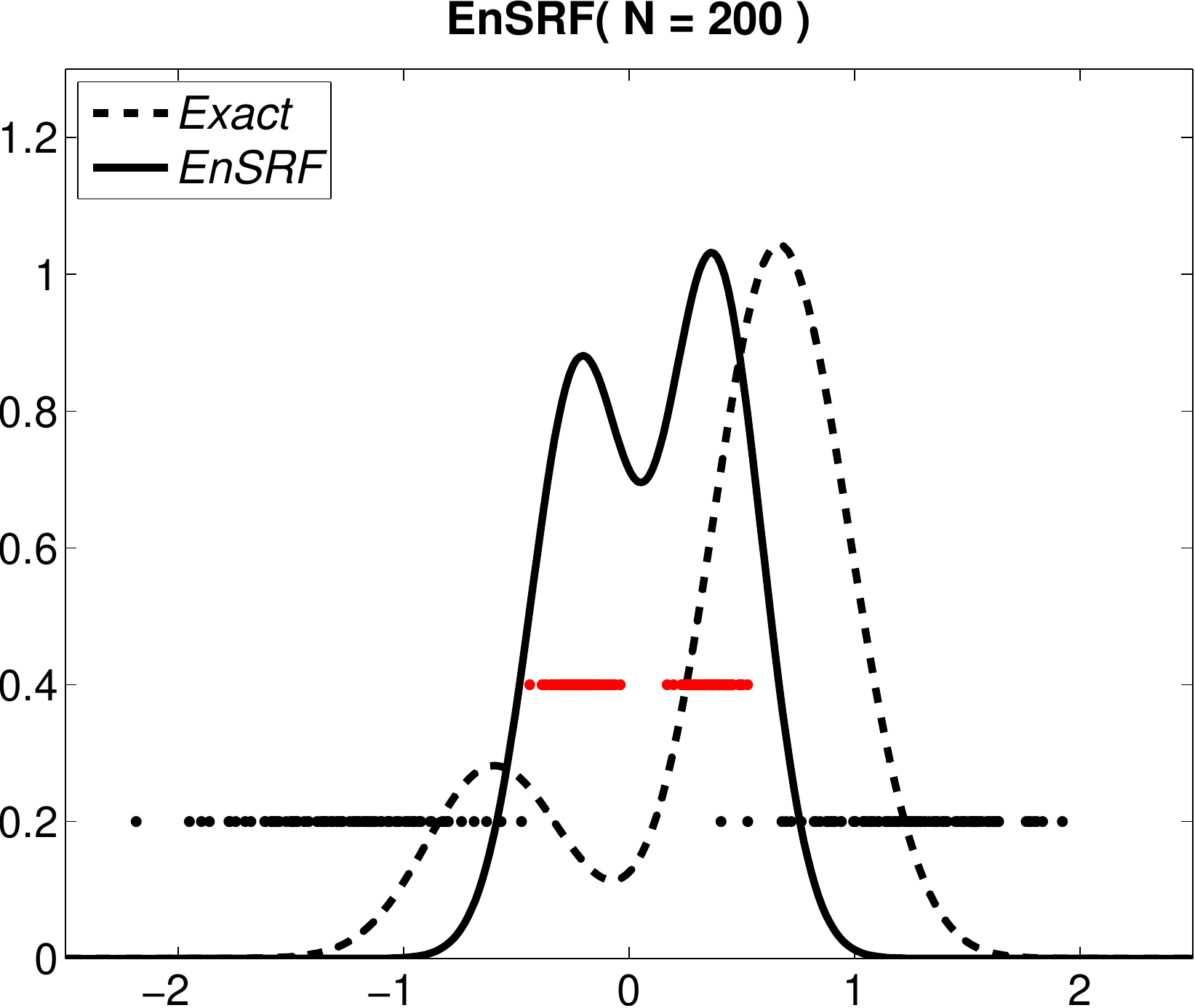}
\end{minipage}
&
\hspace{-25pt}
\begin{minipage}{125pt}
\centering
\includegraphics[width=125pt]{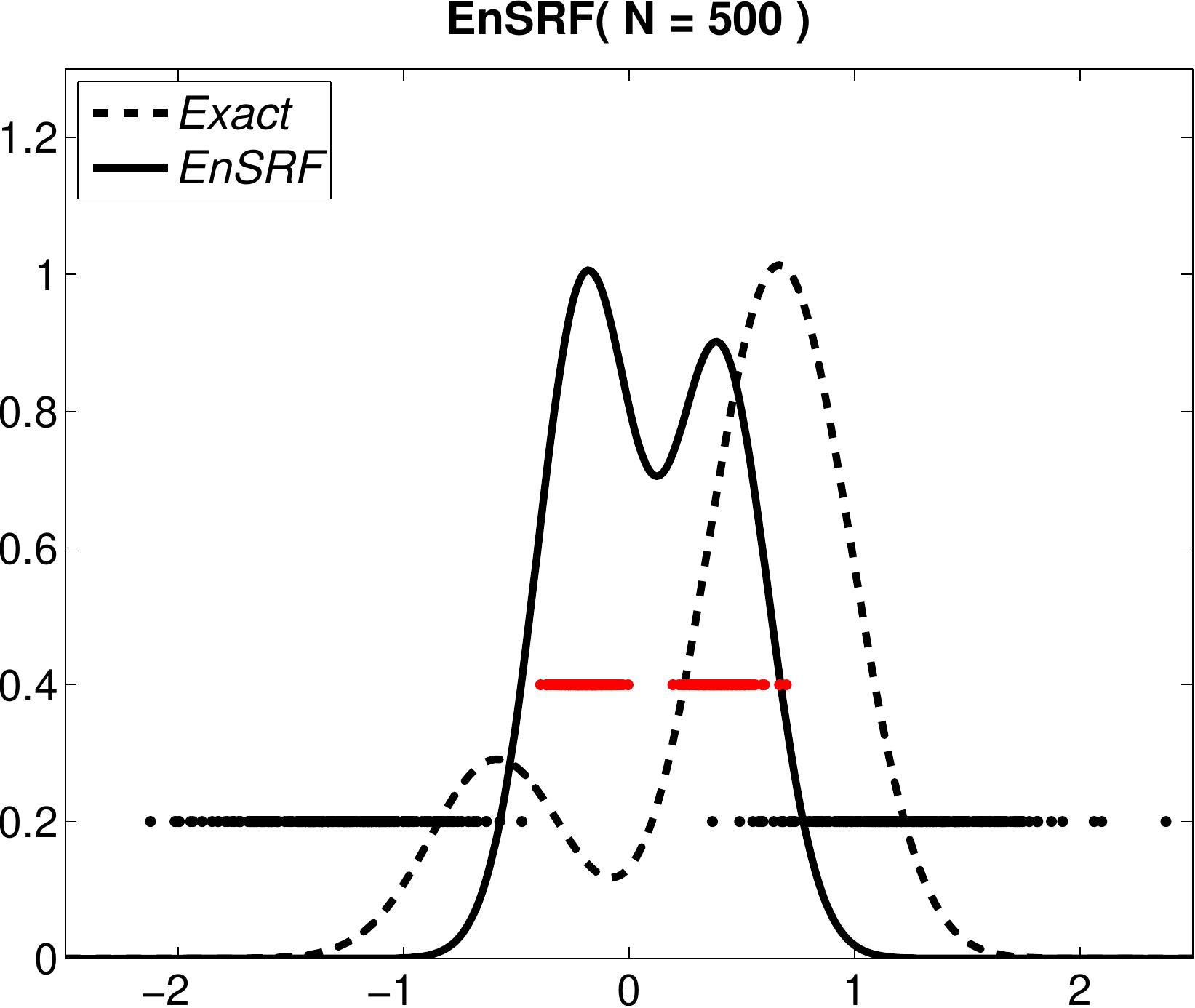}
\end{minipage}
&
\hspace{-25pt}
\begin{minipage}{125pt}
\centering
\includegraphics[width=125pt]{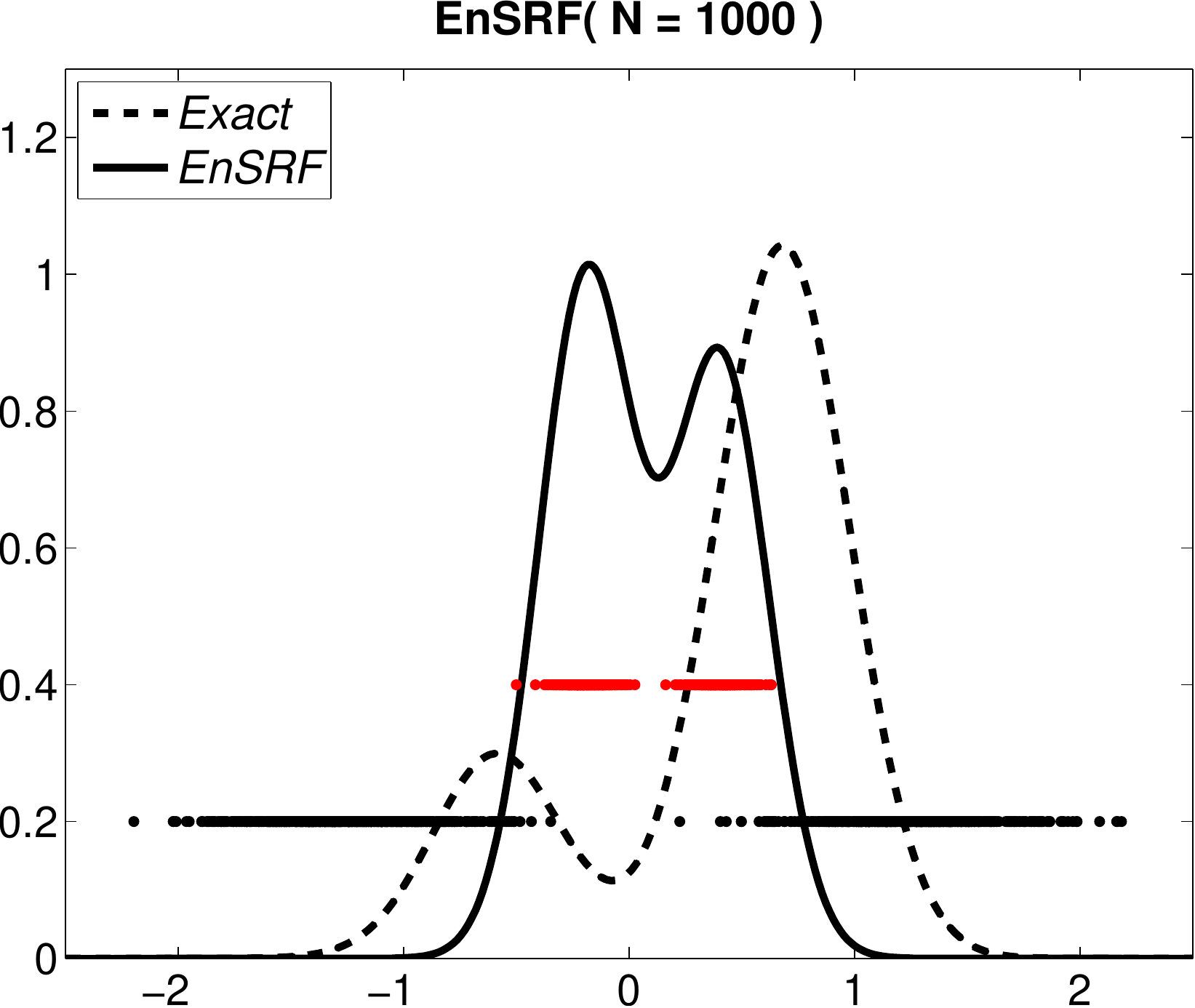}
\end{minipage}

\\
\begin{minipage}{15pt}
\centering
\hspace{-50pt}
EnGSF
\end{minipage}
&
\begin{minipage}{125pt}
\centering
\includegraphics[width=125pt]{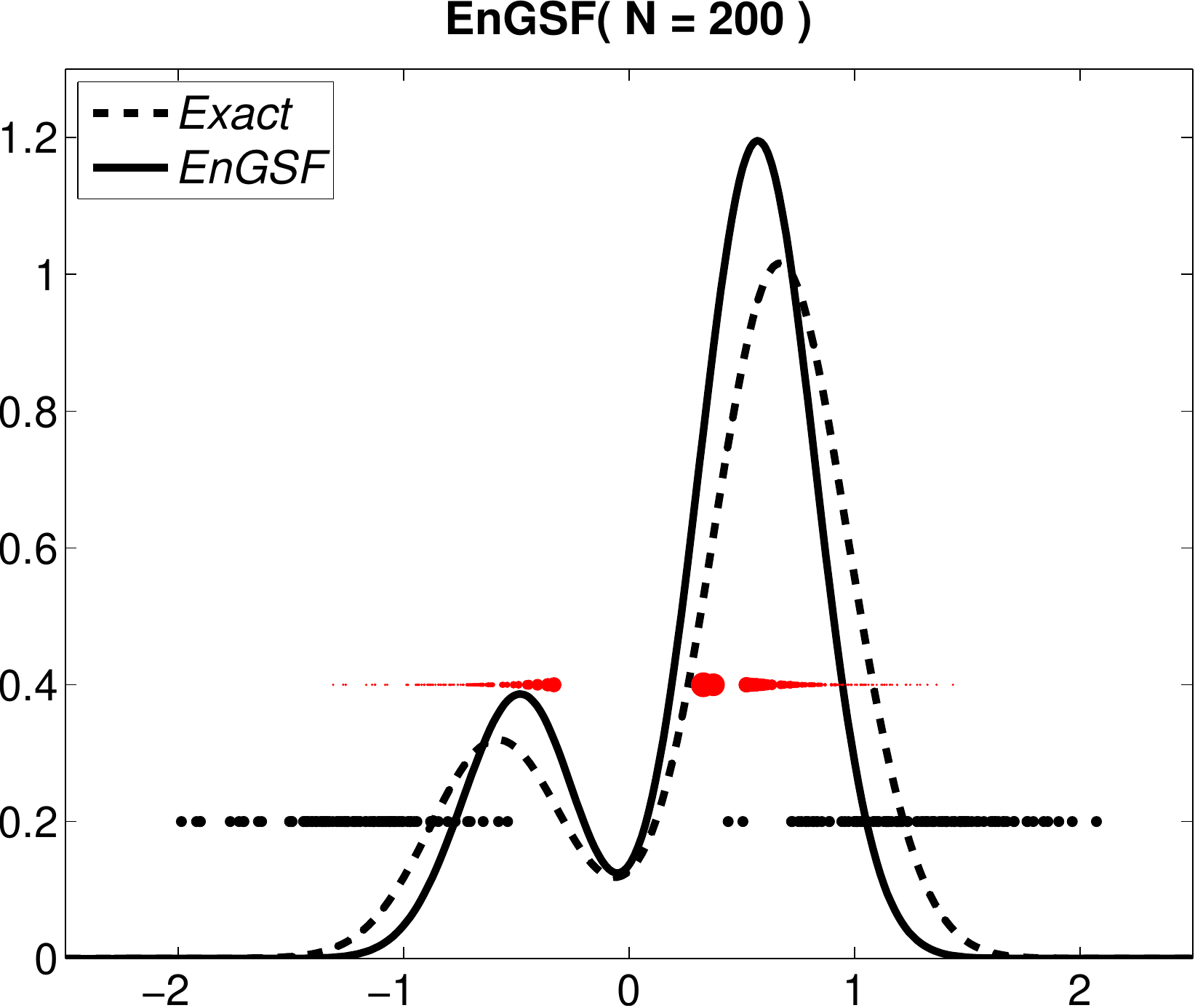}
\end{minipage}
&
\hspace{-25pt}
\begin{minipage}{125pt}
\centering
\includegraphics[width=125pt]{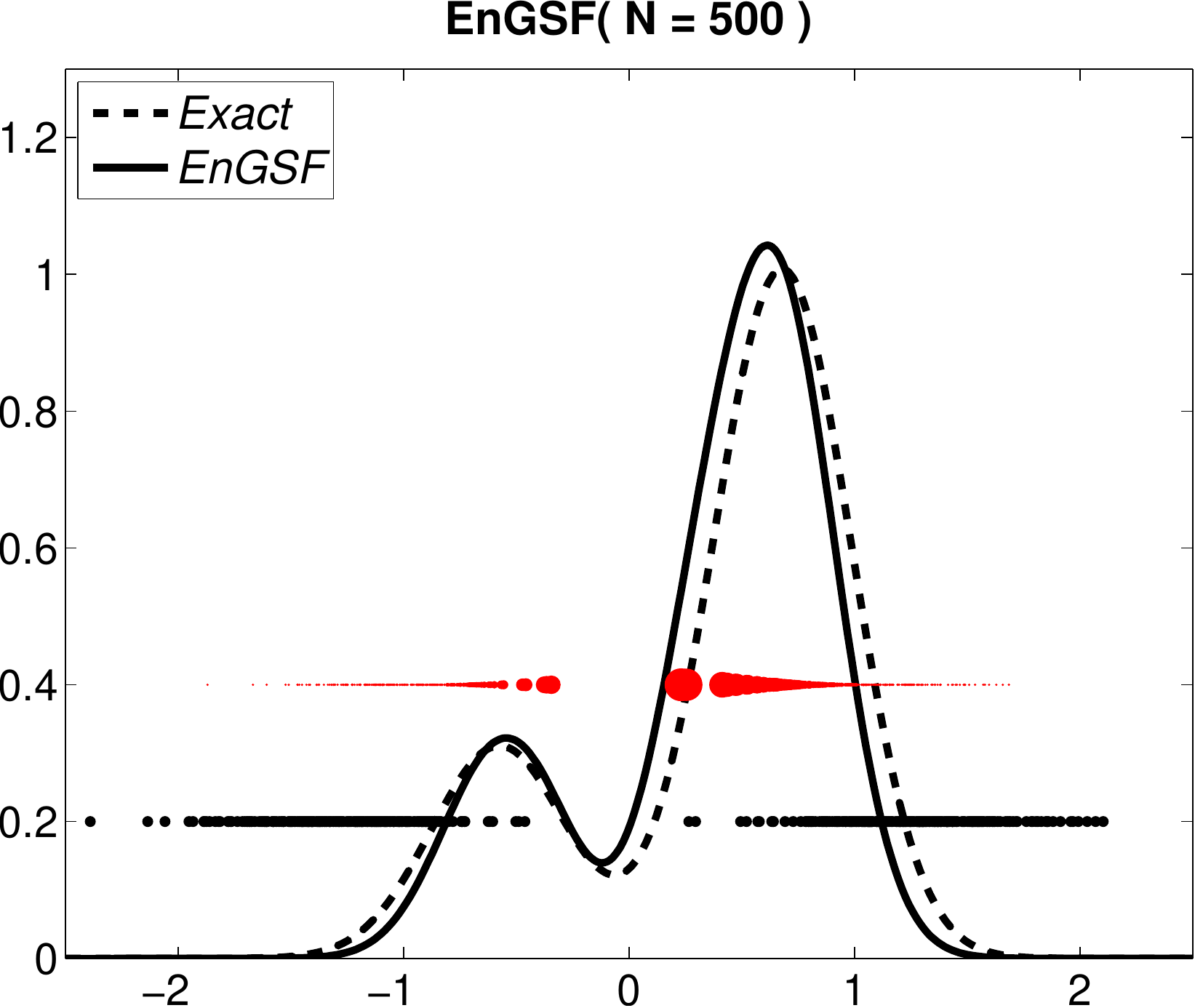}
\end{minipage}
&
\hspace{-25pt}
\begin{minipage}{125pt}
\centering
\includegraphics[width=125pt]{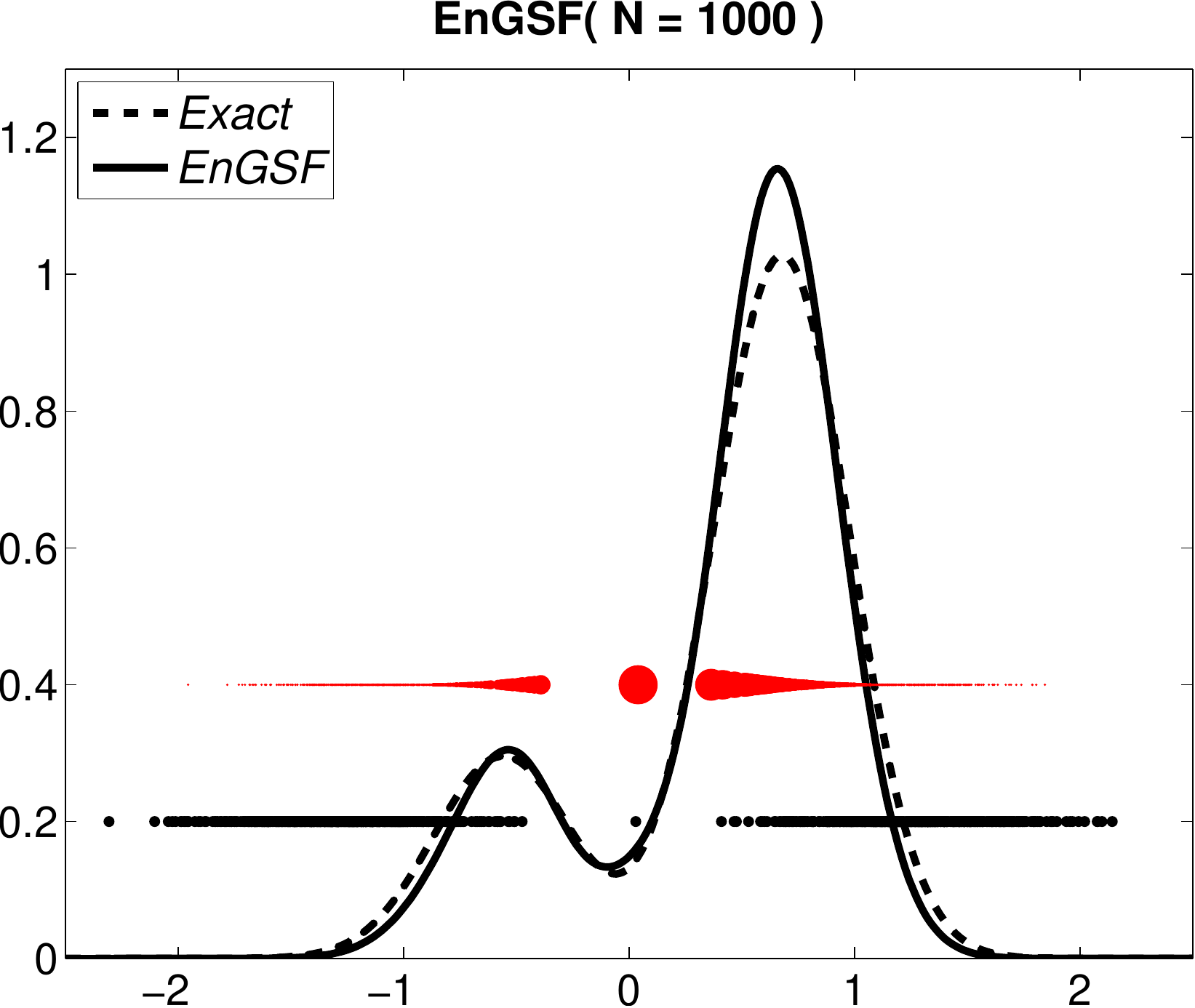}
\end{minipage}

\end{tabular}
\caption{Posterior pdf obtained from different filtering techniques for various number of particles}
\label{fig:apprx_posterior}

\end{figure}

Fig.~ (\ref{fig:apprx_posterior}) shows the posterior pdf approximated using three different data assimilation methods, namely \ac{EnKF}, \ac{EnSRF} and EnGSF. Clearly, increasing the number of particles has little effect on the posterior pdf obtained using \ac{EnKF} and \ac{EnSRF} while that of EnGSF converges to the true posterior. This may be attributed to the fact that EnKF and its variants are optimal w.r.t only the first two statistical moments. Another interesting observation one can make from these plots is the difference between EnKF and EnSRF while both methods are from the same family. It is clear that EnSRF is capable of detecting some bimodal behavior whereas EnKF tend to smooth out bimodality. This discrepancy may be attributed to the fact that in EnKF, observations are treated as random variables and perturbed in order to maintain a reasonable variance. This results show another bad side effect of perturbing observations in addition to those pointed out in \cite{whitaker02}.

The movement of particles are also plotted in the same figure. The bottom and top particles are those of prior and posterior respectively. The size of particles is proportional to their weights. It is clear that the transformation of particles is of significant importance in capturing the accurate posterior \ac{pdf}. Owing to the fact that in EnKF particles are assumed equally weighted, ideally one needs a transformation that takes the i.i.d samples of prior and outputs i.i.d samples of posterior if one wants to achieve the correct pdf. Clearly the transformation performed in the EnKF does not do so for a general pdf. 

A measure of the difference between two \ac{pdf} may be obtained using Kullback-Leibler (KL) divergence. we use this measure in order to study the convergence behavior of different data assimilation methods in this paper. The KL divergence for continuous probability density functions $p$ and $q$ is defined as:
\begin{equation}
D_{KL} (p \| q) = \int_{-\infty} ^{\infty}  \log \frac{p(x)}{q(x)}  \ p(x) \ dx
\label{eq:KL_cont}
\end{equation} 
Here, $p$ is chosen to be the true \ac{pdf} while $q$ is the approximate pdf obtained using one of the data assimilation techniques. A discrete approximation of eq. (\ref{eq:KL_cont}) may be obtained by:

\begin{equation}
\tilde{D}_{KL} (p \| q) = \sum_{i} \log \frac{p(x_i)}{q(x_i)}  \ p(x_i) \ \Delta x_i
\label{eq:KL_disc}
\end{equation}

\begin{figure}[htbp]
\centering
{\includegraphics[width= 200pt,keepaspectratio]{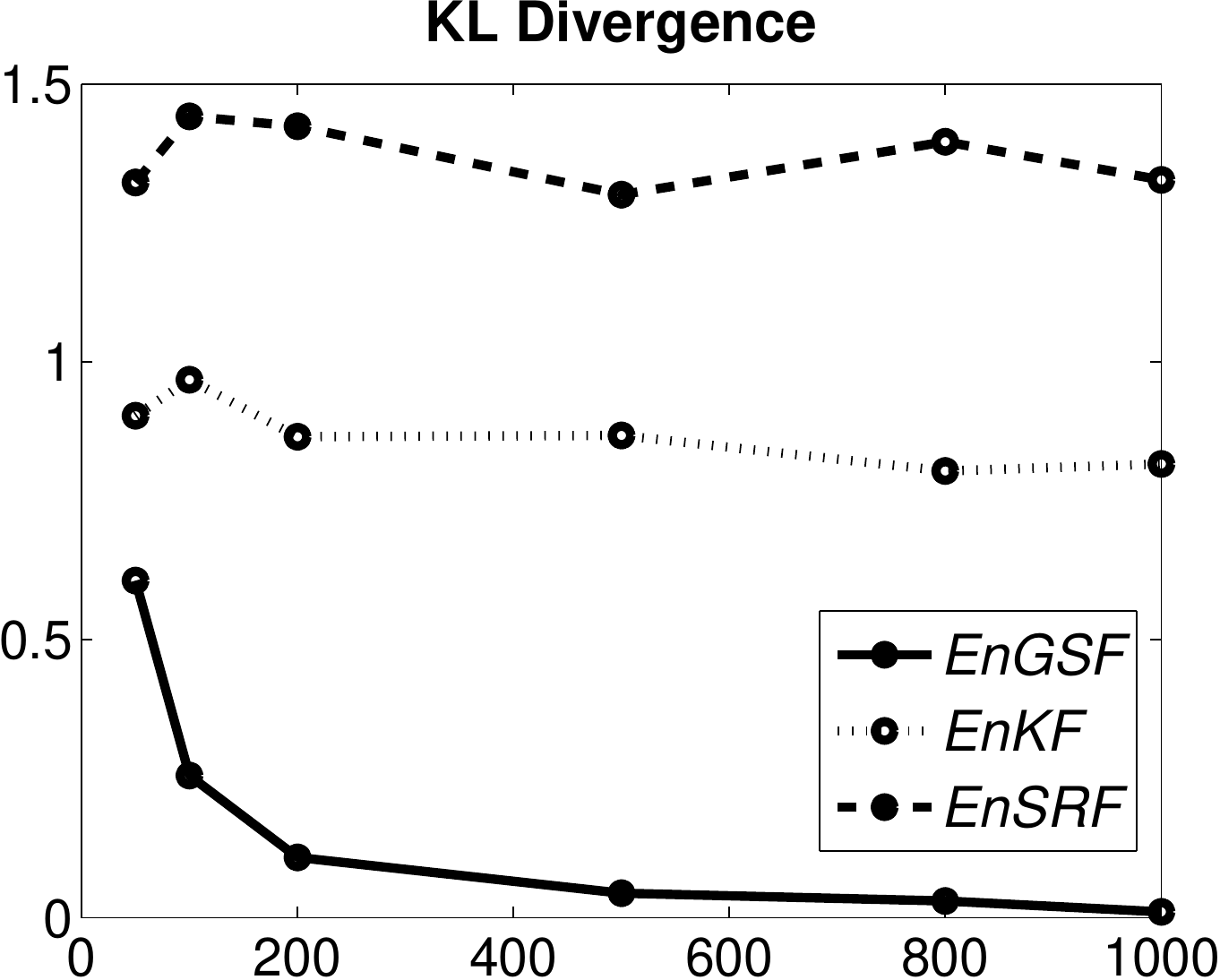}}
\caption{The convergence plot of different methods according to KL divergence measure}
\label{fig:KLD}
\end{figure}

Fig.~ (\ref{fig:KLD}) shows the KL divergence curve for the three methods discussed in this paper. Clearly, the convergence of EnKF and EnSRF is insensitive to the number of particles, $N$, used in the process of data assimilation. EnGSF, however, shows a monotonic convergence as $N$ increases.  

\subsection{Example 2: One-dimensional nonlinear stochastic differential equation}
In this example, we apply the same methods discussed in the previous example and the \ac{SIR} filter to predict the state of a nonlinear stochastic differential equation (SDE) using noisy and sparse observational data. Note that \ac{SIR} filter is the same as \ac{SIS} filter in which one performs resampling after each data assimilation step. We consider the following It\'{o} \ac{SDE}:

\begin{equation}
\frac{\partial u(t)} {\partial t}  = 4 u(t) - 4 u(t)^3 + \kappa \xi (t)
\end{equation} 

Here, $\xi(t)$ is a Gaussian white noise representing the modeling uncertainty and $\kappa$ determines the magnitude of the noise. The deterministic part of the above SDE $f'(u) = 4 u(t) - 4 u(t)^3$ corresponds to the potential function of the form $f(u) = 2u^2 - u^4 $ for which there are three equilibrium points, two stable at $u = \pm1$ and one unstable at $u = 0$.

For $\kappa$ greater than a certain threshold, the state of the above SDE can change from one stable fixed point to the other. Keeping the amount of noise small, however, makes such transitions rare. The pdf of state while system is in transition between stable fixed points is usually non-Gaussian. Therefore, a possible test to assess the performance of a nonlinear filter is to observe its behavior in tracking such transition \cite{mandel09}.

\begin{figure}[htbp]
\centering
{\includegraphics[width= 200pt,keepaspectratio]{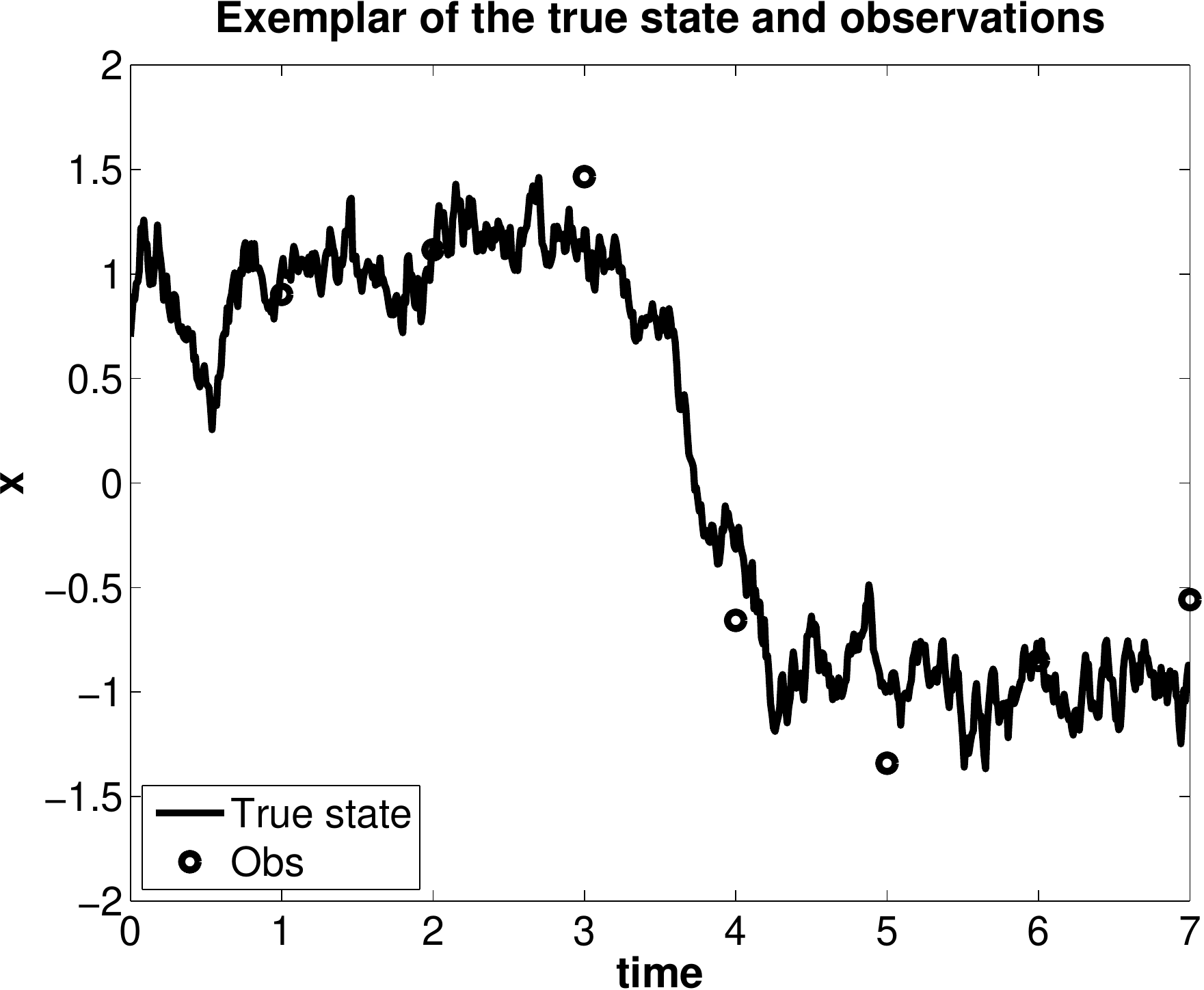}}
\caption{Exemplar of the true state and observations for one dimensional \ac{SDE}}
\label{fig:true_state_SDE}
\end{figure}

Fig.~ (\ref{fig:true_state_SDE}) shows an exemplar of the true state and observations generated using $\kappa = 0.7$, $u(0) = 0.8$ and variance of the noise $R = 0.1$. 

\begin{figure}[ht]
\centering
\begin{tabular}{cc}
\begin{minipage}{150pt}
\centering
\includegraphics[width=150pt]{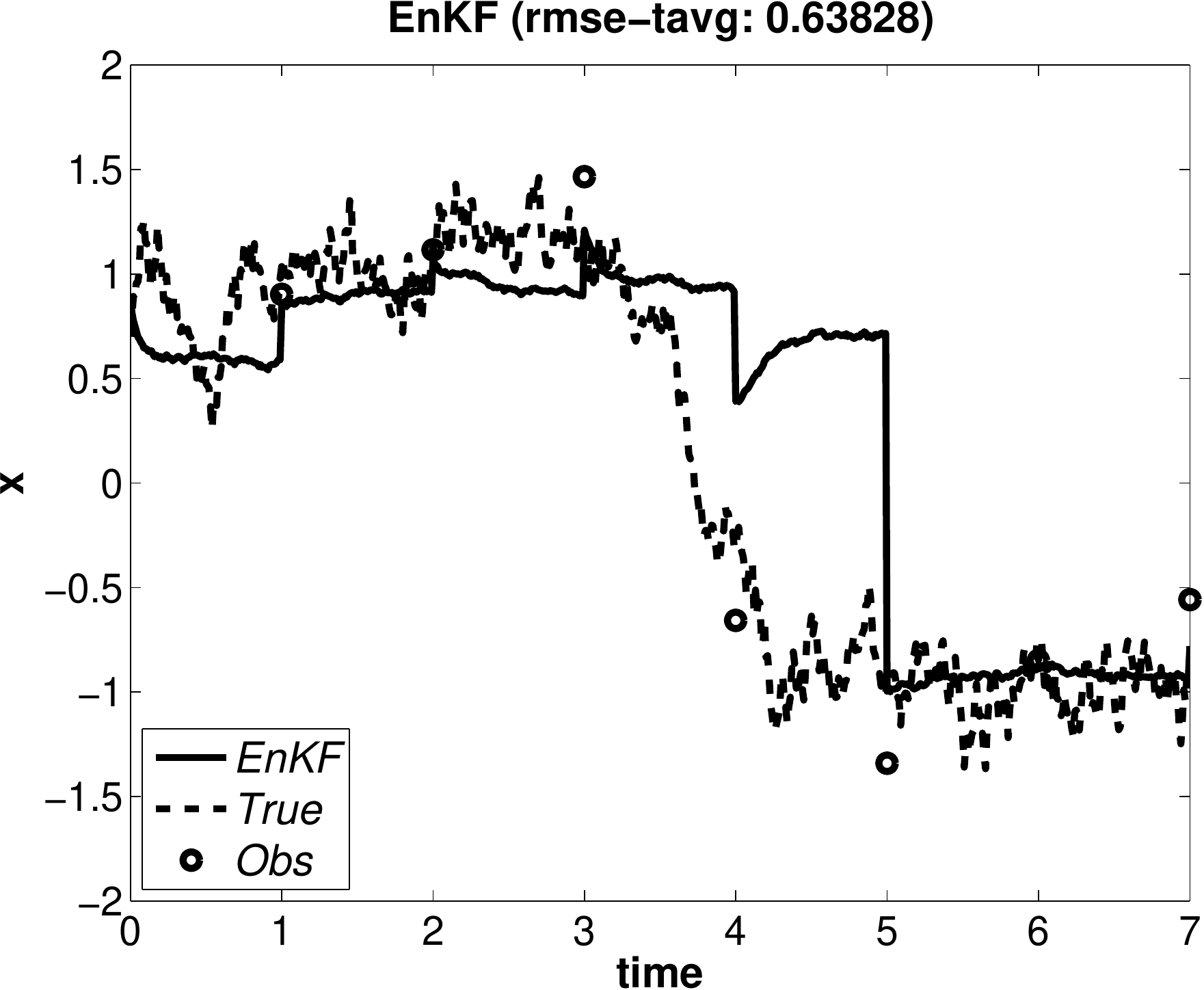}
\end{minipage}
&
\begin{minipage}{150pt}
\centering
\includegraphics[width=150pt]{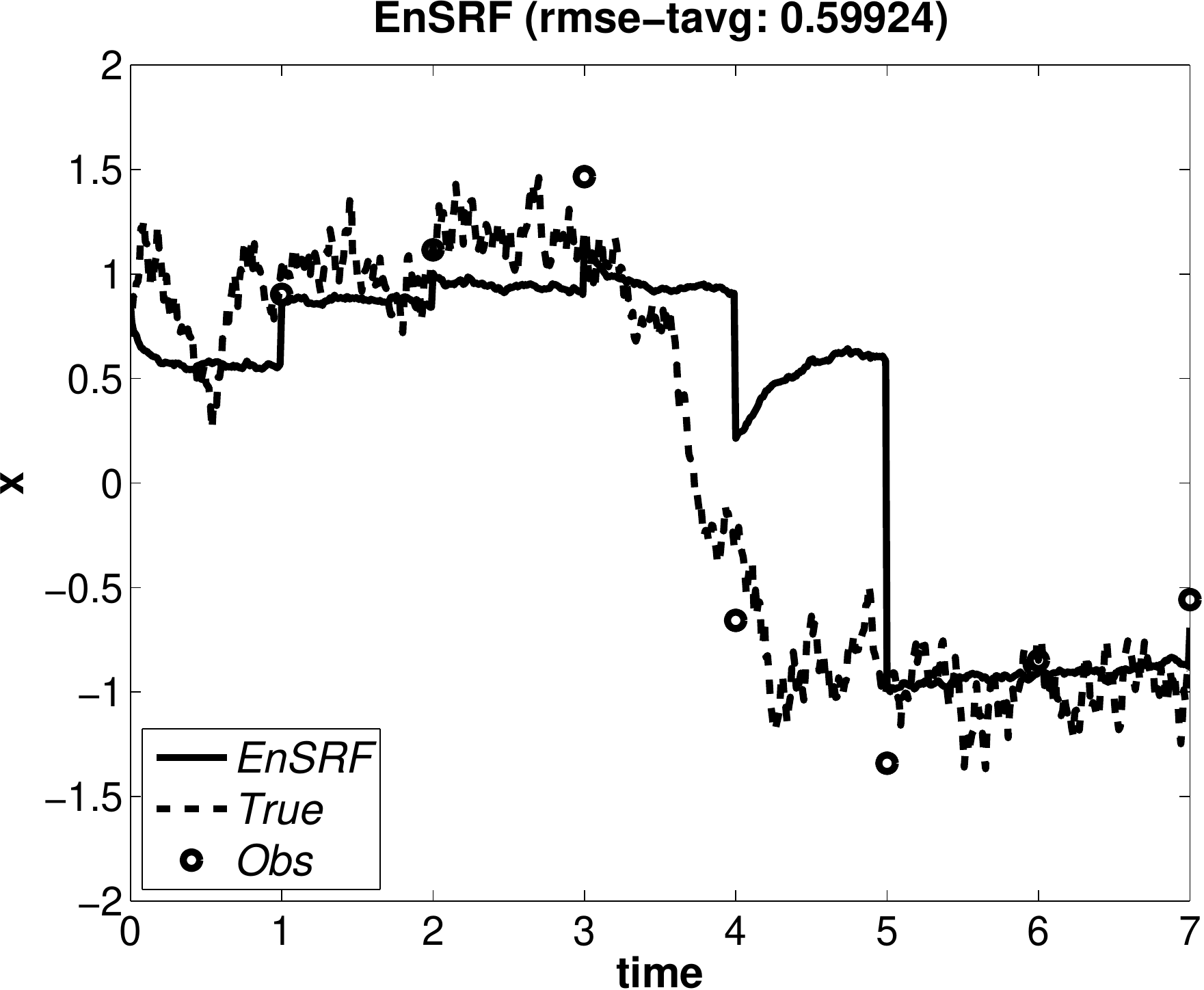}
\end{minipage}
\\
\begin{minipage}{150pt}
\centering
(a)
\end{minipage}
&
\begin{minipage}{150pt}
\centering
(b)
\end{minipage}

\\
\begin{minipage}{150pt}
\centering
\includegraphics[width=150pt]{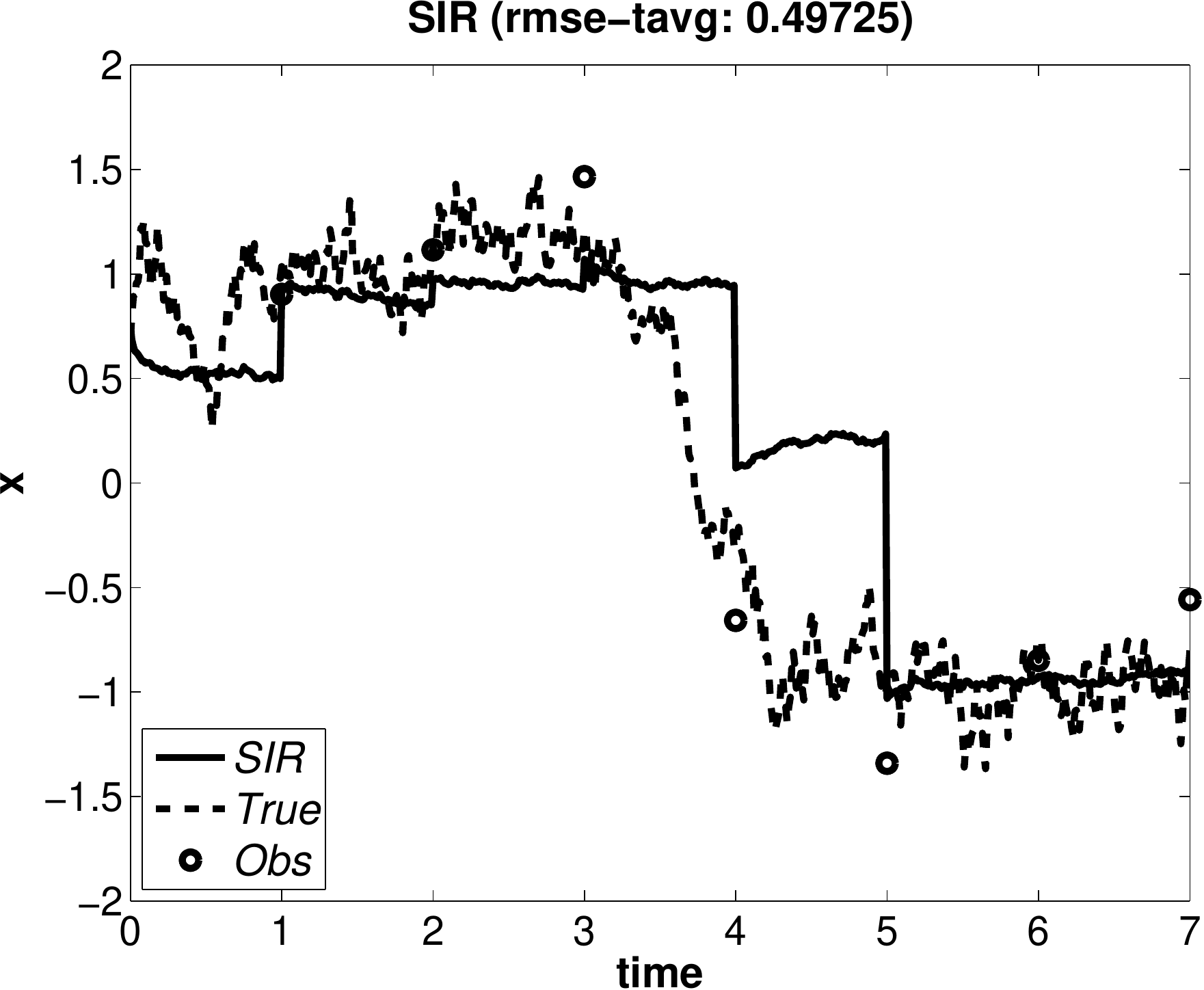}
\end{minipage}
&
\begin{minipage}{150pt}
\centering
\includegraphics[width=150pt]{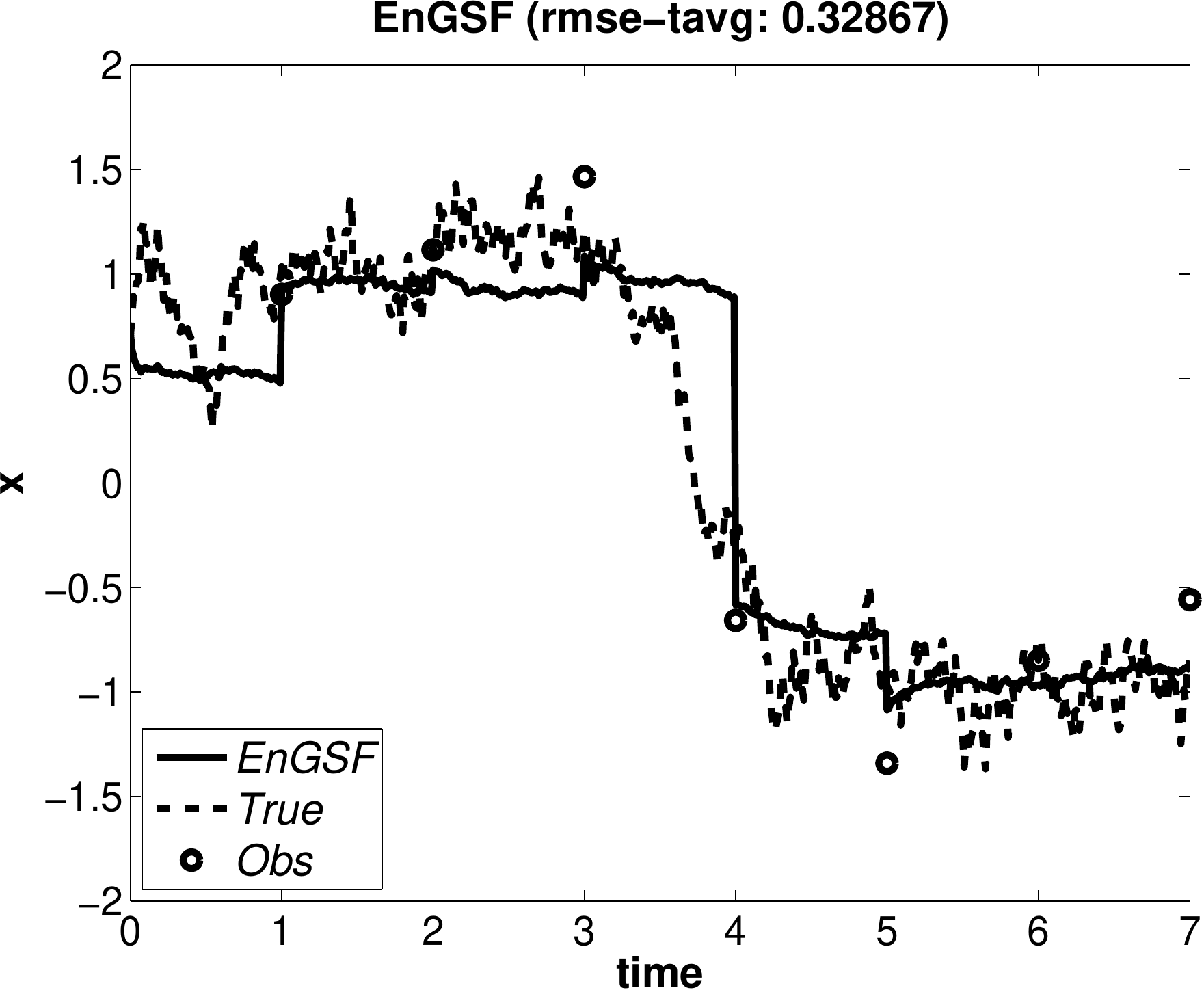}
\end{minipage}
\\
\begin{minipage}{150pt}
\centering
(c)
\end{minipage}
&
\begin{minipage}{150pt}
\centering
(d)
\end{minipage}

\end{tabular}
\caption{Mean estimate obtained from different data assimilation techniques with $N=100$ particles: (a) EnKF (b) EnSRF (c) SIR and (d) EnGSF}
\label{fig:SDEmean}
\end{figure}

Fig.~ (\ref{fig:SDEmean}) shows the estimated mean for different schemes with $N = 100$. As is clear, EnKF and EnSRF are slow in making the transition as also reported in \cite{mandel09}. Also, the difference between the two is minimal as expected in case with larger number of particles. It must be noted that EnSRF may result in a slightly better estimates due to using a modified gain matrix in the update step which eliminates the need of perturbing observations. However, in the experience of the authors its limitation for incorporating measurements serially limits the use of vectorized computer architecture and may result in excessive overhead when the number of data is large. Moreover, the transformation of the gain matrix may not be accurate when covariance localization is applied ans some care must be taken as pointed out in \cite{whitaker02}. 

It is a known fact that SIR filter does not track the transition effectively when the number of particles is small which is consistent with the results obtained here. EnGSF, however, tracks the transition more accurately. This is attributed to the fact that the posterior pdf approximated by EnGSF tend to be more accurate than those of other schemes in areas of non-Gaussian behavior. As a result of this, the time averaged root mean squared error (RMSE) of the estimate obtained by EnGSF is almost half the one of EnKF ($0.33$ vs. $0.64$). Though we have reported the mean here, it must be noted that the mean may not be an appropriate measure to compare different algorithms in case of non-Gaussian/nonlinear problems as also pointed out in \cite{arulampalam02}. 

\begin{figure}[ht]
\centering
\begin{tabular}{cc}

\begin{minipage}{150pt}
\centering
\includegraphics[width=150pt]{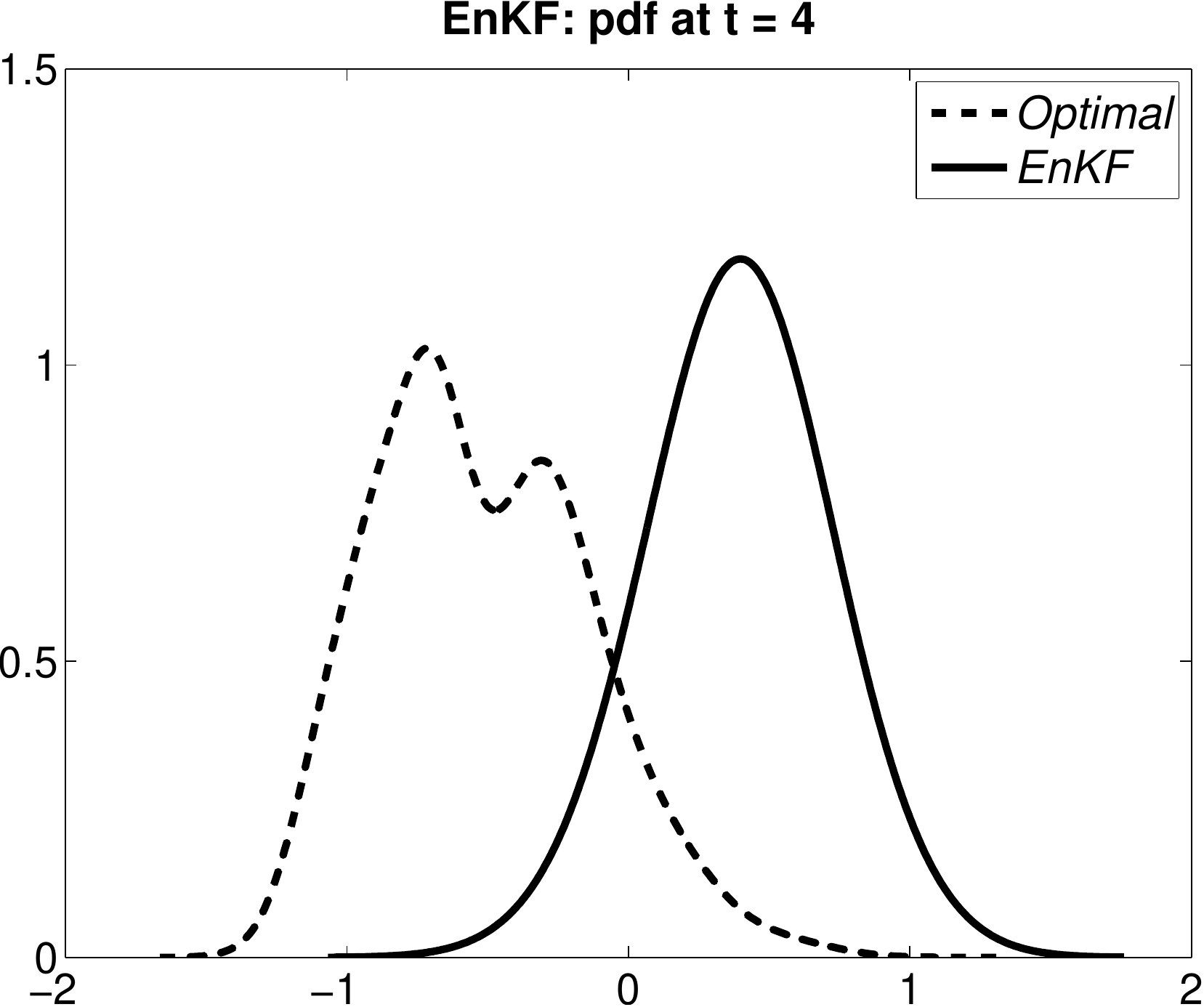}
\end{minipage}
&
\begin{minipage}{150pt}
\centering
\includegraphics[width=150pt]{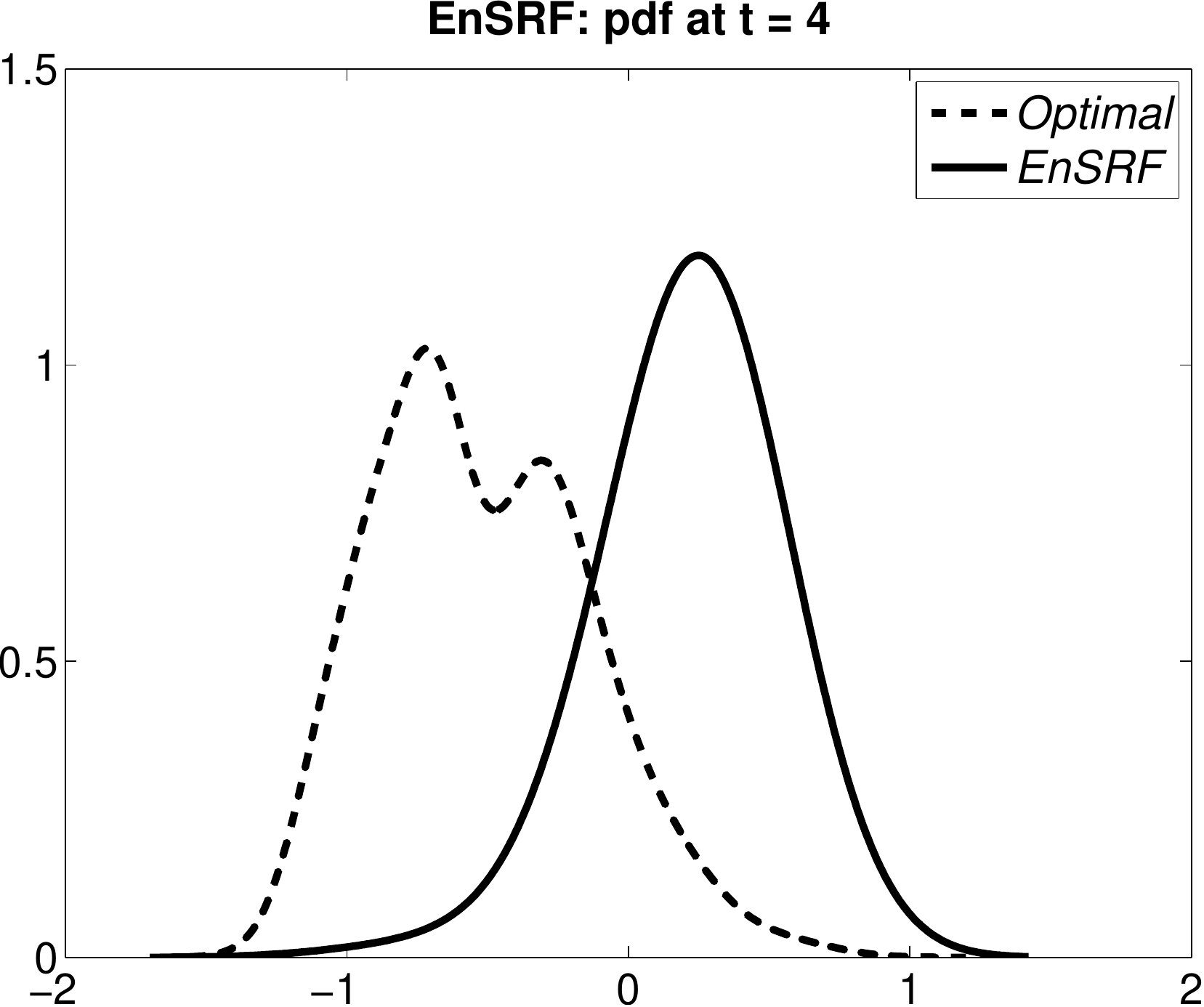}
\end{minipage}
\\
\begin{minipage}{150pt}
\centering
(a)
\end{minipage}
&
\begin{minipage}{150pt}
\centering
(b)
\end{minipage}

\\
\begin{minipage}{150pt}
\centering
\includegraphics[width=150pt]{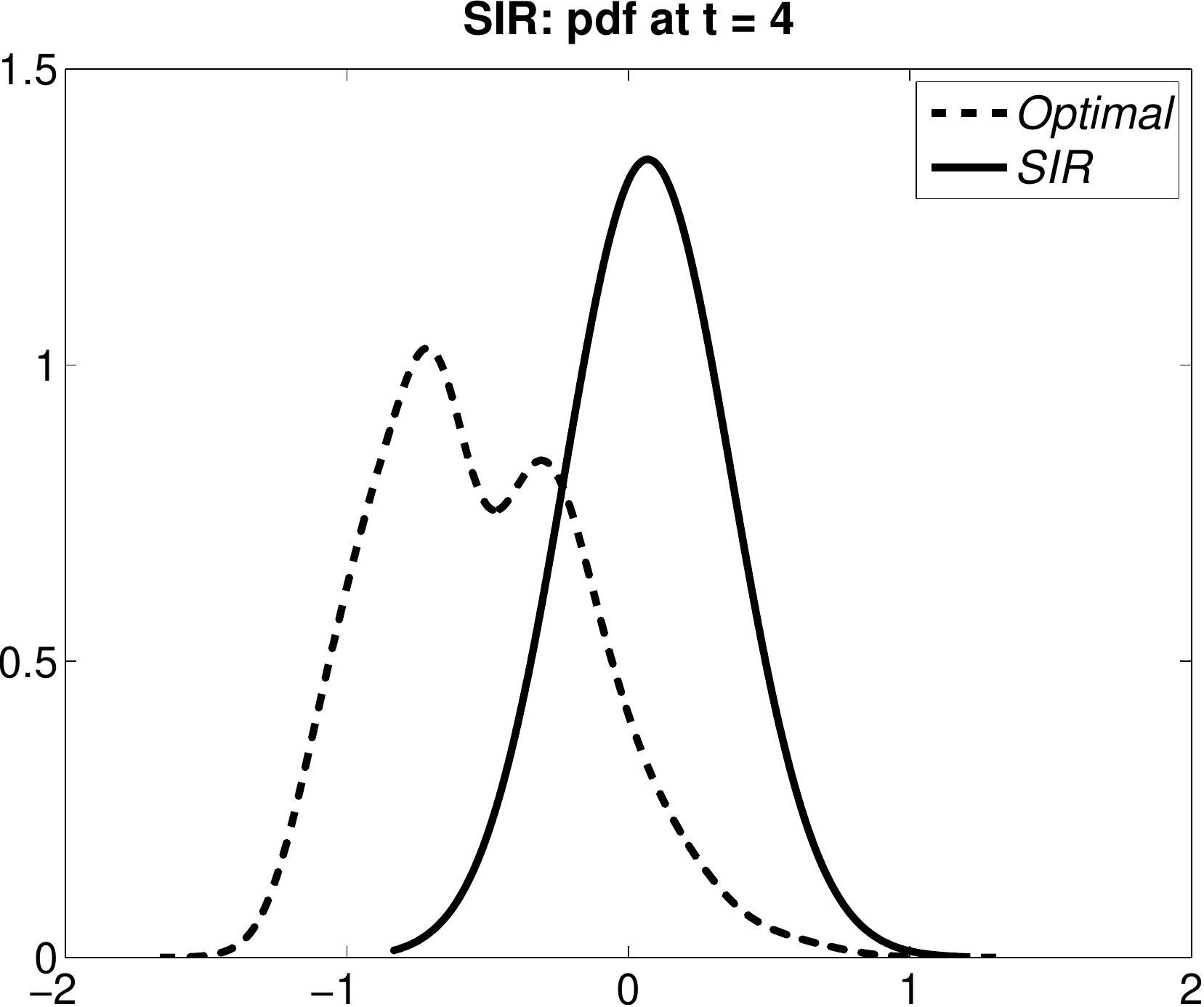}
\end{minipage}
&
\begin{minipage}{150pt}
\centering
\includegraphics[width=150pt]{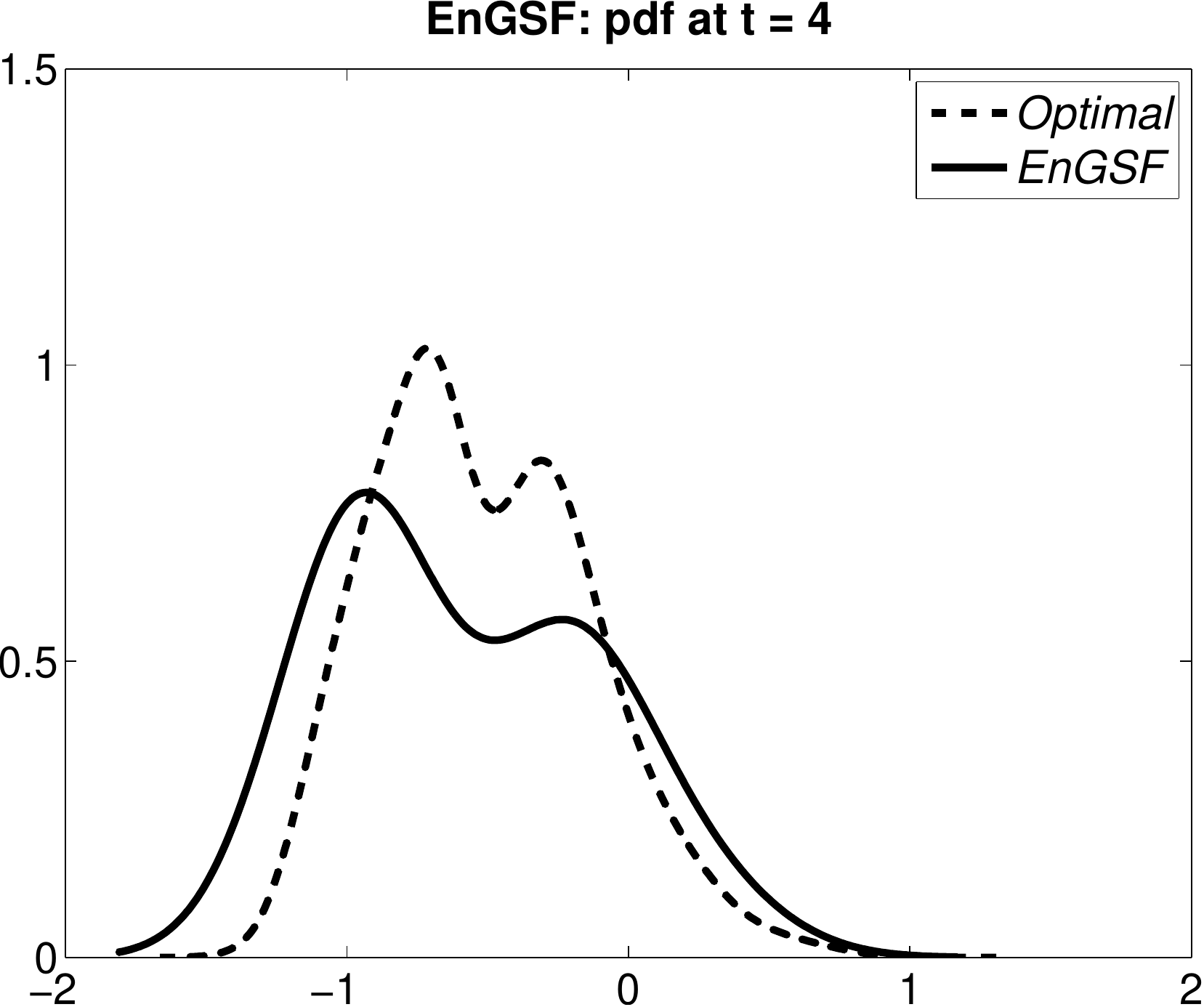}
\end{minipage}
\\
\begin{minipage}{150pt}
\centering
(c)
\end{minipage}
&
\begin{minipage}{150pt}
\centering
(d)
\end{minipage}

\end{tabular}
\caption{Posterior pdf at $t=4$ obtained from different data assimilation techniques with $N=100$ particles: (a) EnKF (b) EnSRF (c) SIR and (d) EnGSF. The optimal posterior pdf is obtained using SIR filter with $N = 10,000$ particles}
\label{fig:SDEpdf}
\end{figure}

Fig.~ (\ref{fig:SDEpdf}) shows the posterior pdf at the time of transition, (i.e., $t = 4$). The optimal posterior density function is obtained using \ac{SIR} filter with $10,000$ particles. While EnKF, EnSRF and SIR produce a biased and unimodal estimate of the posterior, EnGSF results in a more accurate estimate of the posterior pdf which captures the non-Gaussian features effectively.   

\subsection{Example 3: Three dimensional Lorenz63 model}
The Lorenz63 model \cite{lorenz63} is a set of three coupled nonlinear ordinary differential equations defined by:
\begin{eqnarray}
\frac{dx(t)}{dt} = \gamma (y-x), \ \frac{dy(t)}{dt} = \rho x - y - xz , \  \frac{dz(t)}{dt} = xy - \beta z
\end{eqnarray}  
where x(t), y(t), z(t) are time dependent unknown variables. The commonly chosen parameters are $\gamma =10$, $\rho = 28$ and $\beta = 8/3$. To integrate the above equations, we use a forth order Runge-Kutta scheme with a time step of $\Delta t = 0.01$.  
The Lorenz63 model has been the subject of many data assimilation studies due to its nonlinear and chaotic nature (c.f., \cite{evensen99,nakano07, whitaker02}). Here, we study the behavior of EnGSF and EnKF in tracking the state of the lorenz63 model.

A reference solution is generated by integrating the model in the time interval $t \in [0, 100]$ with the initial condition $ (x_0; y_0; z_0) = (1.508870; -1.531271; 25.46091)$ and a diagonal model error covariance with variances equal to 2.0, 12.13 and 12.31 corresponding to the error growth over one time unit for $x$, $y$ and $z$ respectively. These are the same values used in reference \cite{evensen99}. The stochastic term is applied as $dq = \sigma \sqrt{\Delta t} \ d\xi $ where $\sigma^2$ is the variance and $d\xi$ is drawn from unit normal. Observations are collected for all variables by adding Gaussian noise with standard deviation of $2.5$ to the reference solution and the distance between consecutive observations are assumed $\Delta t_{obs} = 0.5$. At this level of sparsity and noise, it is expected that non-Gaussian features become important.  
For the data assimilation, we used the same modeling error covariance by which the reference solution is obtained. The initial realizations are generated from a multivariate normal distribution having mean equal to $(x_0; y_0; z_0)$ and covariance equal to $ 4 \bI_3$. The number of particles is set to $N = 200$

\begin{figure}[htp]
\centering
\begin{tabular}{c}
\begin{minipage}{400pt}
\centering
\includegraphics[width=400pt, height = 200 pt]{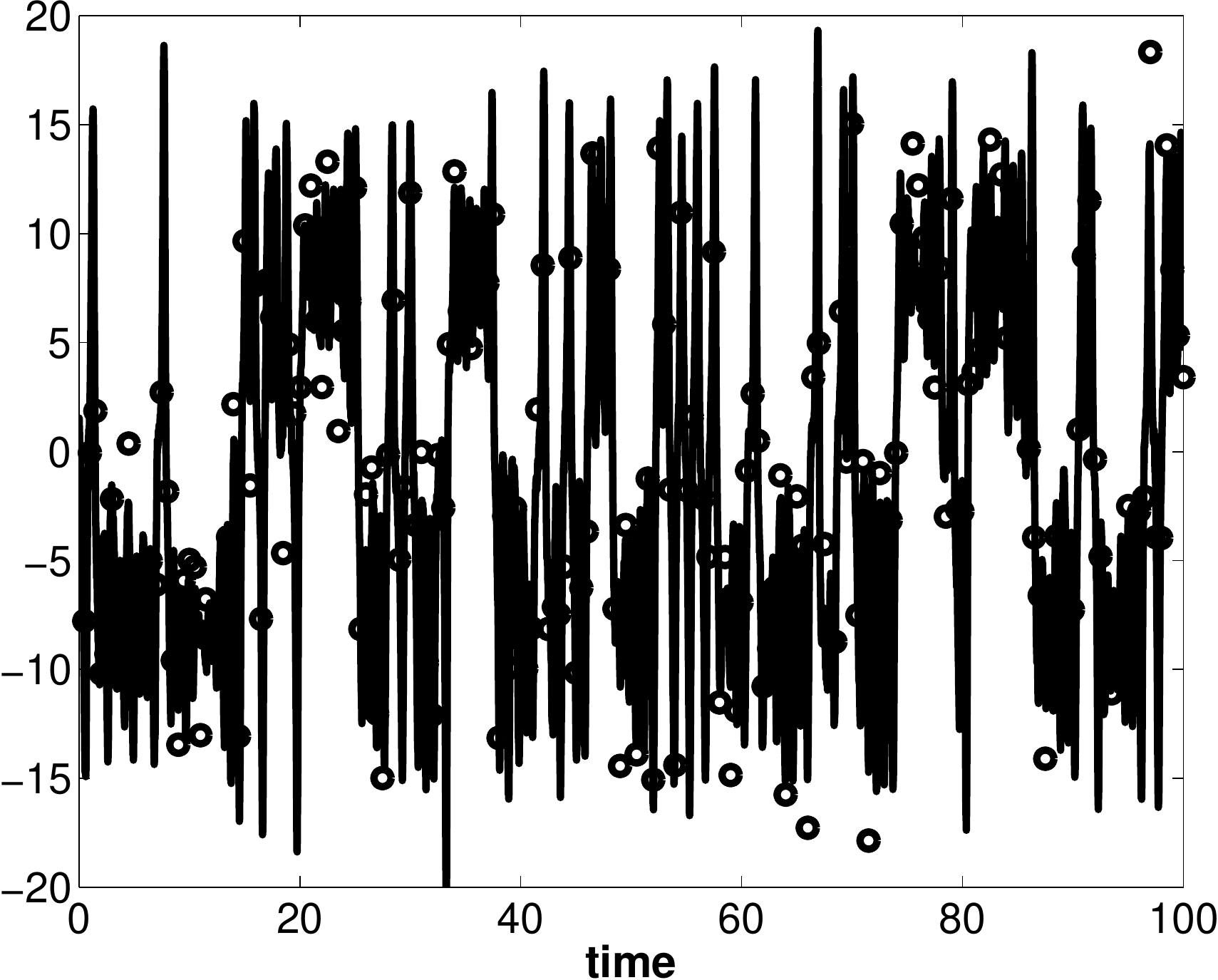}
\end{minipage}
\\
\begin{minipage}{400pt}
\centering
\includegraphics[width=400pt, height = 200 pt]{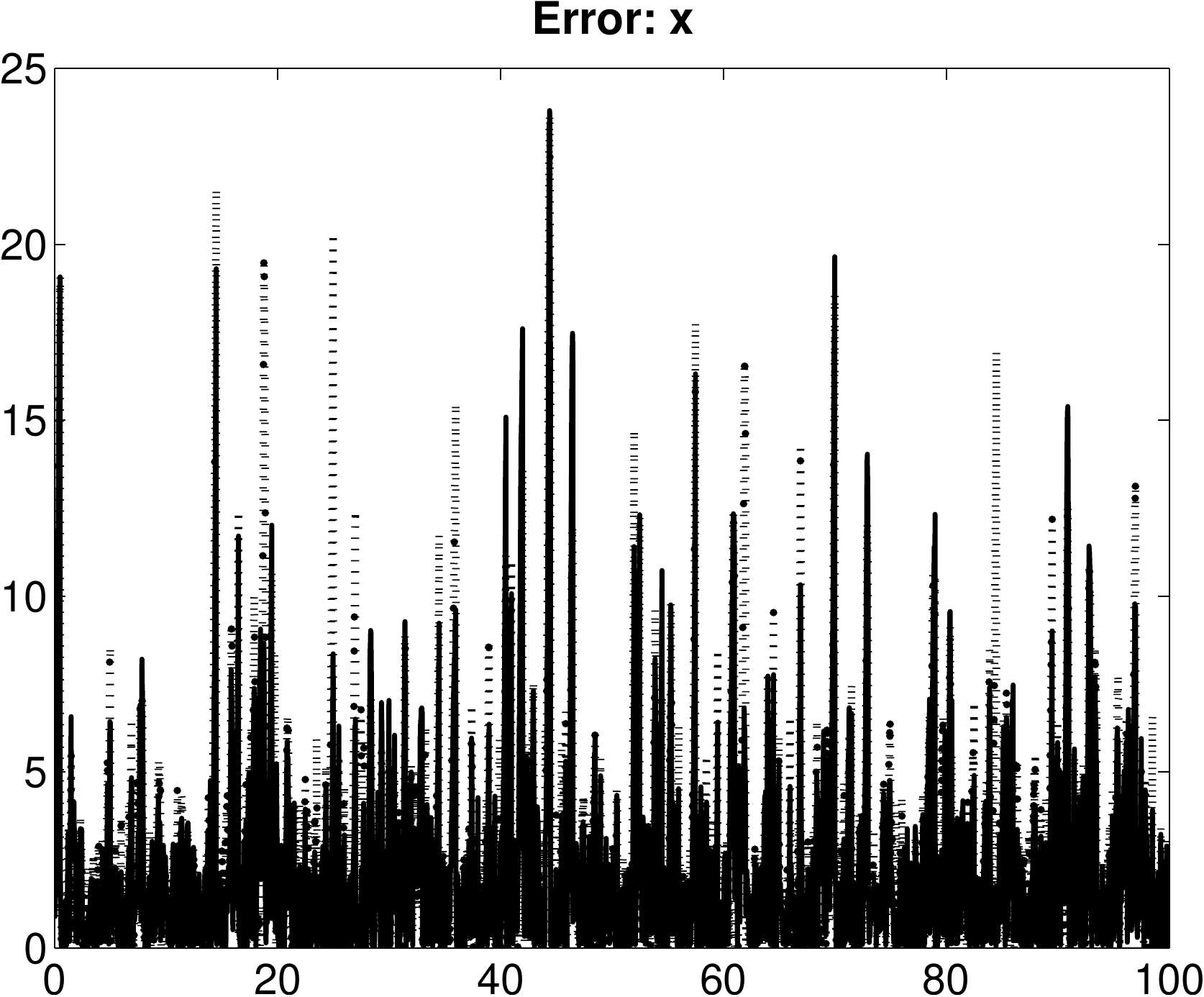}
\end{minipage}
\\
\end{tabular}

\label{fig:Lorenzx}
\caption{(top) The reference solution for x. (bottom) The absolute value of the difference between the reference solution and the estimates obtained by EnKF (dotted line) and EnGSF (solid line)}
\end{figure}

\begin{figure}[htp]
\centering
\begin{tabular}{c}
\begin{minipage}{400pt}
\centering
\includegraphics[width=400pt, height = 200 pt]{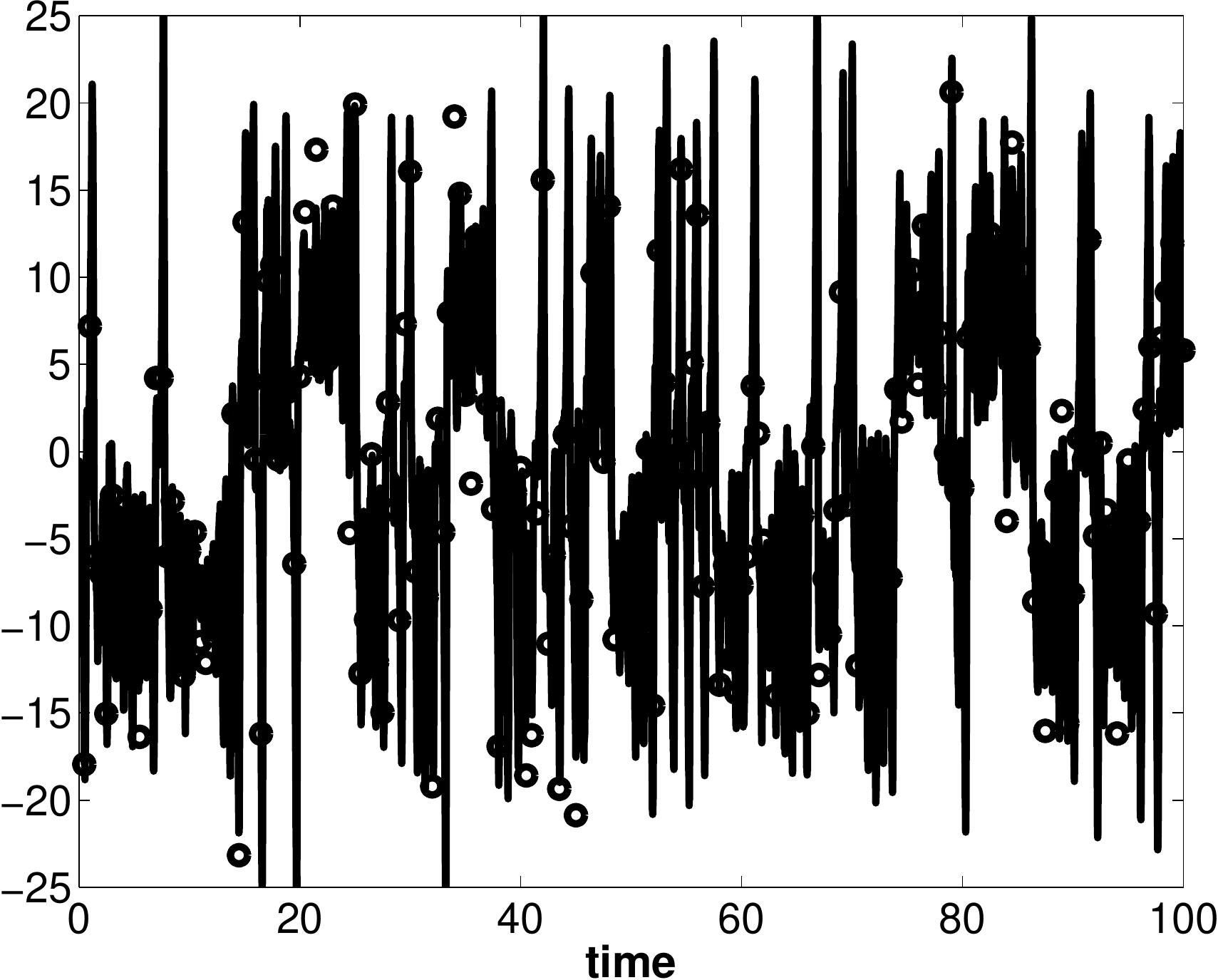}
\end{minipage}
\\
\begin{minipage}{400pt}
\centering
\includegraphics[width=400pt, height = 200 pt]{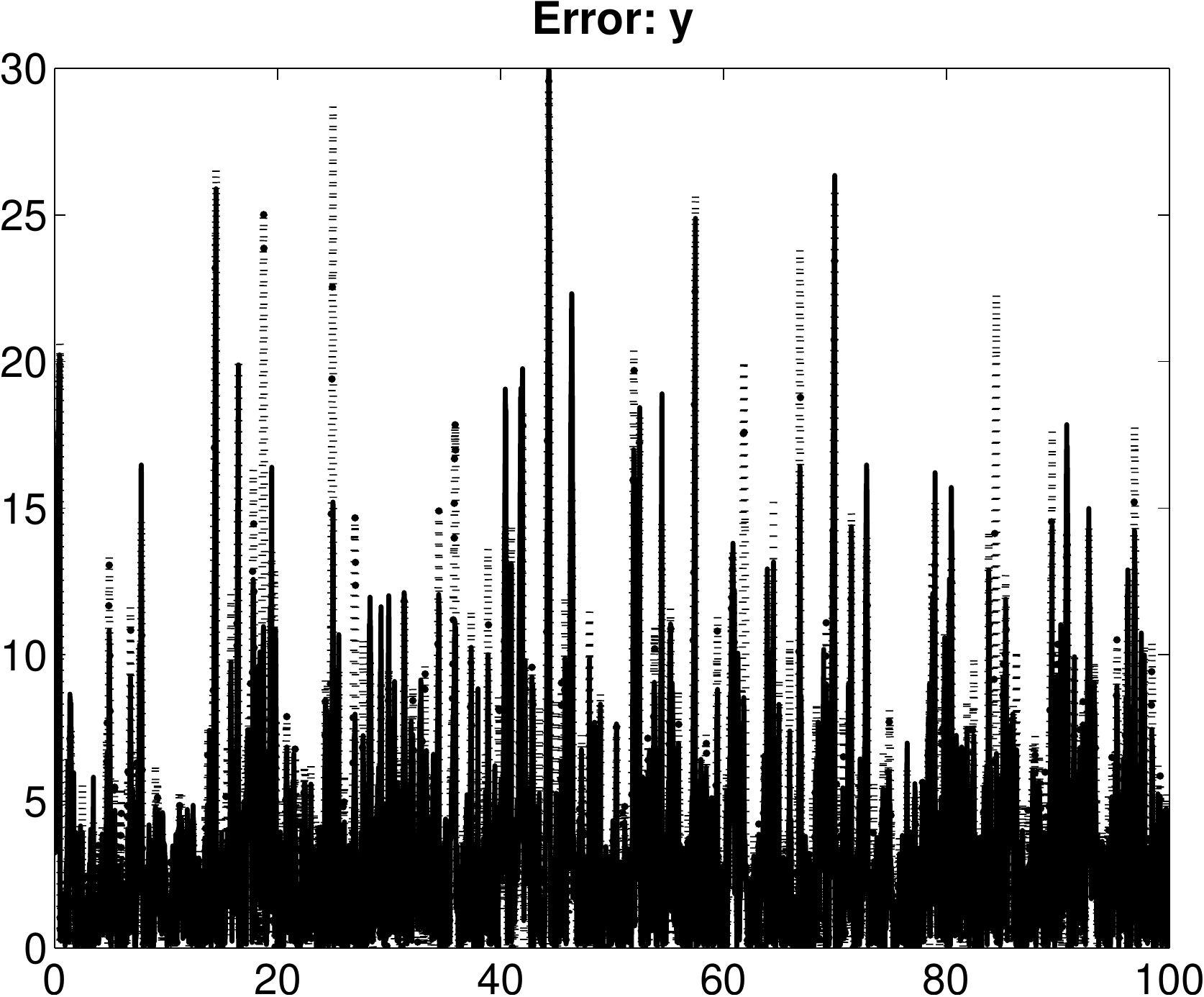}
\end{minipage}
\\
\end{tabular}

\label{fig:Lorenzy}
\caption{(top) The reference solution for y. (bottom) The absolute value of the difference between the reference solution and the estimates obtained by EnKF (dotted line) and EnGSF (solid line)}
\end{figure}

Fig. (\ref{fig:Lorenzx}) show the reference solution for $x$ and the absolute value of error between the reference solution and the estimates obtained using EnKF and EnGSF. Similar plot is shown in Fig. (\ref{fig:Lorenzy}) for the $y$ variable. Clearly, large errors take place in the transition areas (such as $t=25$) as also pointed out in \cite{evensen99}. However, we observe that EnGSF results in lower errors compraed to that of EnKF in such areas. For this experiment, we observed a time-averaged RMSE of $3.42$ for EnGSF and $3.74$ for EnKF. It is notable that this number for SIR filter with 2000 particles was $3.39$. 

\subsection{Example 4: Forty dimensional Lorenz95 model}
Lorenz95 model \cite{lorenz95} is a system of coupled nonlinear ordinary differential equations defined by:
\begin{equation}
\frac{dx_{j}}{dt}  = (x_{j+1} - x_{j-2}) x_{j-1} - x_{j} + F, \ \ \ j = 1 \cdots m
\end{equation} 
with periodic boundary conditions defined by $x_{-1} = x_{m-1}$ , $x_{0} = x_{m}$ and $x_{1} = x_{m+1}$. This model resembles dynamics of the atmosphere and has been shown to behave chaotically for $F$ greater than $4.0$ and m = 40 \cite{lorenz98}. Though the dimension of state space for this model is still far from those of the real world, It is considered as a challenging problem from the perspective of data assimilation because of its highly chaotic nature \cite{vanleeuwen10}.  

Here we set $m = 40$, $F= 8.0 $ and the initial condition is chosen to be $(x_j= F, \ j\neq 20)$ and $x_{20} = 1.001 F$. A forth order Runge-Kutta scheme with a time step of $\Delta t = 0.01$ is used to integrate the model. It is shown in \cite{lorenz98} that a time step of 0.05 unit in the model corresponds to 6-h in real life. Observations are collected for all variables every $5$ time step ($6$-h) for a period of 5000 time step after a spinup period of 2000 time steps. Note that this choice of observational time corresponds to a regime where the effect of non-Gaussian statistics are significant \cite{vanleeuwen10, nakano07}. The observational noise is considered Gaussian according to $N(\textbf{0}, 2 \bI_{40})$. The initial prior pdf is considered Gaussian with uncorrelated variables each having mean and variance equal to $2.0$. The model noise is applied as $dq = \sigma \sqrt{\Delta t} \ d\xi $ where the variance $\sigma^2 $ is set to 25.0 and $d\xi$ is drawn from unit normal.

\begin{figure}[htbp]
\centering
{\includegraphics[width= 250pt,keepaspectratio]{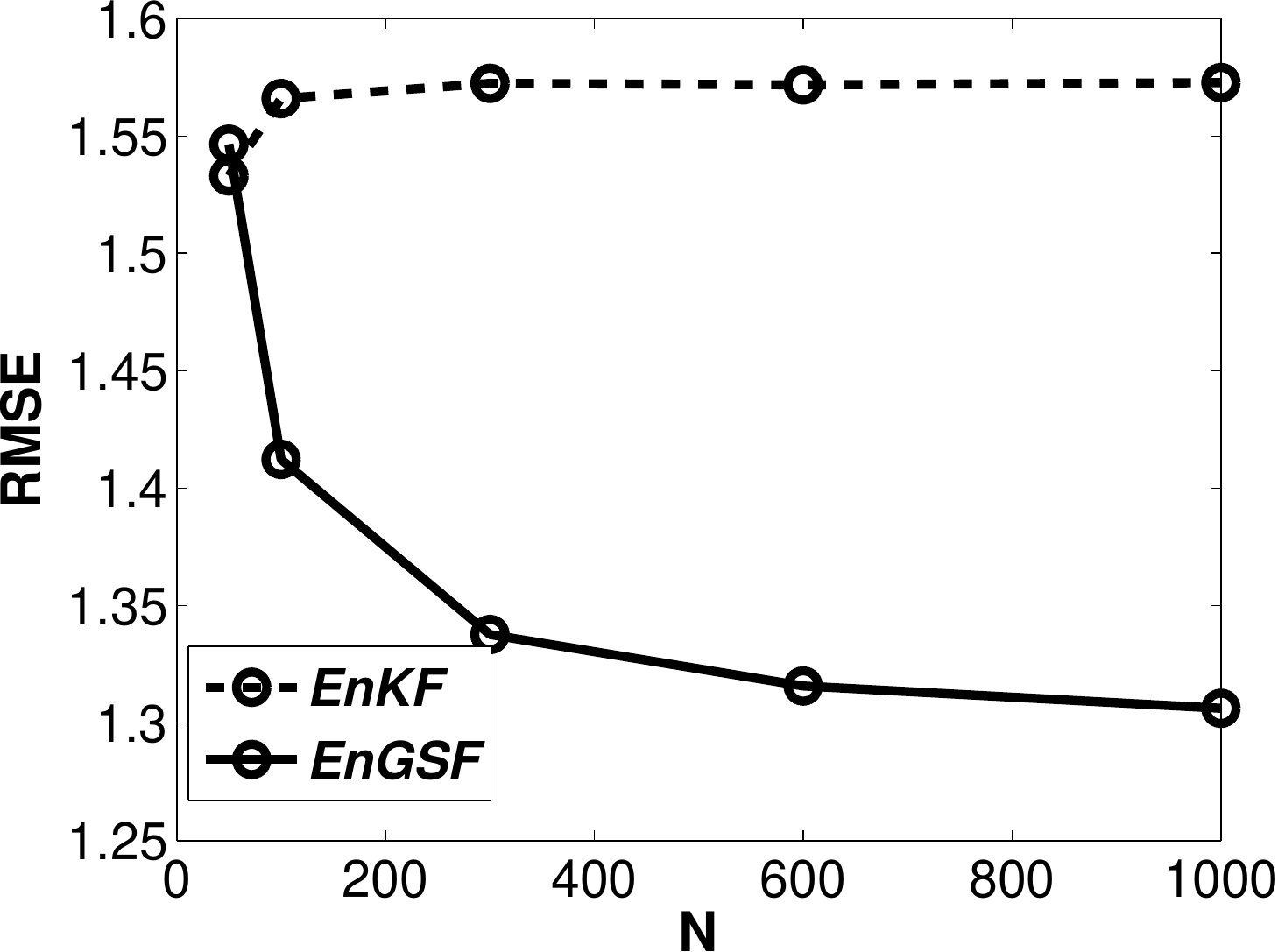}}
\caption{Time averaged \ac{RMSE} vs. number of particles for Lorenz95 model using EnGSF and EnKF}
\label{fig:L95RMSE}
\end{figure} 

Fig.~ (\ref{fig:L95RMSE}) shows the time-averaged \ac{RMSE} obtained for Lorenz95 model using EnGSF and EnKF with variable number of particles. Clearly, the EnKF estimate does not improve as the number of particles increases, a phenomenon observed in previous examples too. EnGSF, on the other hand shows a monotonic convergence with the number of particles and produces better estimates compared to EnKF in this setting.

\section{Conclusion} \label{sec:concl}

We have presented a new methodology, the \ac{EnGSF}, for nonlinear tracking and estimation problem. The \ac{EnGSF}, as shown through several numerical examples, has the ability to represent non-Gaussian statistics more accurately than currently operational data assimilation schemes such as \ac{EnKF}. This may be beneficial for dynamics with highly non-linear behavior. 

Moreover, as computers become more powerful, one would expect that the number of forward runs one can perform in a certain time interval increases and so data assimilation can be carried out with more particles. This brings about the need for data assimilation schemes that could effectively utilize higher order statistical moments than only the first two. The convergence results of the \ac{EnGSF} presented in this paper show that the method scales well with the number of particles as opposed to EnKF and its variants that are optimal with respect to only the first two moments. 
  
Another advantage of the \ac{EnGSF} is its affordable computational cost compared to other nonlinear estimation techniques. For instance, in compound methods that use the posterior pdf obtained by \ac{EnKF} as the proposal density for SIS filter (cf. \cite{mandel09, vanleeuwen10}), the computational cost while calculating the weights may become cumbersome for large-scale problems, though these methods are certainly attractive in that a more accurate representation of pdf is obtained as a result of two stage filtering. Note that the computational cost of \ac{EnGSF} is proportional to that of \ac{EnKF}.

The \ac{EnGSF} has clear connections to SIS filter and hence one expects that new resampling ideas such as those introduced in \cite{nakano07} enhance the performance of the method though not explored in this paper. On the other hand, the methodology can be looked as a weighted collection of EnKF's acting together. Therefore, all the implementation aspects of the \ac{EnKF}, such as localization, are also applicable for the \ac{EnGSF}.

\section{Acknowledgments}
\noindent This research was funded by NSF Award 0934596, CMG Collaborative Research: Subsurface Imaging and Uncertainty Quantification. This work benefited from helpful discussions with or reviews by David L. Donoho, Ram Rajagopal, Craig Bishop and Milad Mortazavi. We are indebted to them.  


{\bibliographystyle{plain}
\bibliography{ref}}

\end{document}